\documentclass[letterpaper,10pt]{IEEEtran}

\newtheorem{theorem}{Theorem}[section]
\newtheorem{lemma}[theorem]{Lemma}

\newtheorem{remark}[theorem]{Remark}
\newtheorem{conject}[theorem]{Conjecture}
\newtheorem{corollary}[theorem]{Corollary}

\usepackage[dvips]{graphicx}
\usepackage{url}
\usepackage{multirow}
\usepackage{cite}
\usepackage{amssymb}
\usepackage{amsmath}
\usepackage{subfigure}
\usepackage{graphicx}
\usepackage{tikz}
\makeatletter

\providecommand{\LyX}{L\kern-.1667em\lower.25em\hbox{Y}\kern-.125emX\@}
\providecommand{\boldsymbol}[1]{\mbox{\boldmath $#1$}}

\providecommand{\tabularnewline}{\\}

\begin{document}


\title{Analysis of Verification-based Decoding on the $q$-ary
Symmetric Channel for Large $q$}




\author{Fan Zhang and Henry D. Pfister \\
 Department of Electrical and Computer Engineering, Texas A\&M
University \\
 \{fanzhang,hpfister\}@tamu.edu %
\thanks{This material is based upon work supported by the National Science Foundation under Grant No. 0747470.
Any opinions, findings, conclusions, or recommendations expressed in this material are those of the authors and do not necessarily reflect the views of the National Science Foundation.}}

\maketitle
\begin{abstract}
A new verification-based message-passing decoder
for low-density parity-check (LDPC) codes is introduced and analyzed for the $q$-ary symmetric
channel ($q$-SC). Rather than passing messages consisting of symbol
probabilities, this decoder passes lists of possible symbols and marks
some lists as verified. The density evolution (DE) equations for
this decoder are derived and used to compute decoding thresholds.
If the maximum list size is unbounded, then one finds that any capacity-achieving
LDPC code for the binary erasure channel can be used to achieve capacity
on the $q$-SC for large $q$. The decoding thresholds are also computed
via DE for the case where each list is truncated to satisfy a maximum
list-size constraint. Simulation results are also presented to confirm
the DE results.
During the simulations, we observed differences between
two verification-based decoding algorithms, introduced by Luby and Mitzenmacher,
that were implicitly assumed to be identical. In this paper,
the node-based algorithms are evaluated via analysis and simulation.


The probability of false verification (FV) is also considered and
techniques are discussed to mitigate the FV. Optimization of the degree
distribution is also used to improve the threshold for a fixed maximum
list size. Finally, the proposed algorithm is compared with a variety
of other algorithms using both density evolution thresholds and simulation
results. 
\end{abstract}

\begin{IEEEkeywords}
low density parity check codes, message passing decoding,
q-ary symmetric channel, verification decoding, list decoding,
false verification.
\end{IEEEkeywords}

\section{Introduction}

Low-density parity-check (LDPC) codes are linear codes that were introduced
by Gallager in 1962 \cite{my_ref:r1} and re-discovered by MacKay
in 1995 \cite{mackay1995good}. The ensemble of LDPC codes that we consider
(e.g. see \cite{my_ref:lubyca1} and \cite{my_ref:urbanke}) is defined
by the edge degree distribution (d.d.) functions $\lambda(x)=\sum_{k\geq2}\lambda_{k}x^{k-1}$
and $\rho(x)=\sum_{k\geq2}\rho_{k}x^{k-1}$. The standard encoding
and decoding algorithms are based on the bit-level operations. However,
when applied to the transmission of data packets, it is natural to
perform the encoding and decoding algorithm at the packet level rather
than the bit level. For example, if we are going to transmit 32 bits
as a packet, then we can use error-correcting codes over the, rather
large, alphabet with $2^{32}$ elements.

Let the r.v.s $X$ and $Y$ be the input and output, respectively, of a $q$-ary symmetric channel ($q$-SC) whose transition probabilities are \[
\Pr (Y=y|X=x)=\left\{ \begin{array}{lcl}
1-p & \mbox{if $x=y$}\\
p/(q-1) & \mbox{if $x\neq y$,}\end{array}\right.\]
where $x,y\in GF(q)$.  The capacity of the $q$-SC, $1+(1-p)\log_{q}(1-p)+p\log_{q}p-p\log_{q}(q-1)$,
is approximately equal to $1-p$ symbols per channel use for large $q$. This implies
the number of symbols which can be reliably transmitted per channel use of the $q$-SC with large $q$ is approximately equal
to that of the BEC with erasure probability $p$. Moreover,
the behavior of the $q$-SC with large $q$ is similar to the BEC
in the sense that: i) incorrectly received symbols from the $q$-SC
provide almost no information about the transmitted symbol and ii)
error detection (e.g., a CRC) can be added to each symbol with negligible
overhead \cite{my_ref:ShokITW04}.

Binary LDPC codes for the $q$-SC with moderate $q$ are proposed and optimized
based on EXIT charts in \cite{my_ref:isit2008weidmann} and \cite{my_ref:LW_turbo08}.
It is known that the complexity of the FFT-based belief-propagation
algorithm, for $q$-ary LDPC codes, scales like $O(q\log q)$. Even
for moderate sizes of $q$, such as $q=256$, this renders such algorithms
impractical. However, when $q$ is large, an interesting
effect can be used to facilitate decoding: if a symbol is received
in error, then it is essentially a randomly chosen element of the
alphabet and it is very unlikely that the parity-check equations
involving this symbol are satisfied.

Based on this idea, Luby and Mitzenmacher develop an
elegant algorithm for decoding LDPC codes on the $q$-SC for large
$q$ \cite{my_ref:luby_and_mitz}. However, their paper did not present simulation results and
left capacity-achieving ensembles as an interesting open problem.
Metzner presented similar ideas earlier in \cite{my_ref:Metz} and
\cite{my_ref:Metz2}, but the focus and analysis is quite different.
Davey and MacKay also develop and analyze a symbol-level message-passing
decoder over small finite fields in \cite{my_ref:Davey}. A number of
approaches to the $q$-SC (for large $q$) based on interleaved Reed-Solomon
codes are also possible \cite{my_ref:ShokITW04} \cite{my_ref:BKY03}.
In \cite{my_ref:isit2004wangshok}, Shokrollahi and Wang discuss two
ways of approaching capacity. The first uses a two-stage approach
where the first stage uses a Tornado code and verification decoding.
The second is, in fact, equivalent to one of the decoders we discuss
in this paper.%
\footnote{The description of the second method in \cite{my_ref:isit2004wangshok}
is very brief and we believe its capacity-achieving nature deserves
further attention.%
} When we discovered this, the authors were kind enough to send us
an extended abstract \cite{my_ref:Shokwangpersonal} which contains
more details. Still, the authors did not consider the theoretical
performance with a maximum list-size constraint, the actual performance
of the decoder via simulation, or false verification (FV) due to cycles
in the decoding graph. In this paper, we describe the algorithm in
detail and consider those details.

Inspired by \cite{my_ref:luby_and_mitz}, we introduce 
list-message-passing (LMP) decoding with verification for LDPC codes
on the $q$-SC. Instead of passing a single value between symbol and
check nodes, we pass a list of candidates to improve the decoding
threshold. This modification also increases the probability of FV.
So, we analyze the causes of FV and discuss techniques to mitigate
FV. It is worth noting that the LMP decoder we consider is somewhat
different than the list extension suggested in \cite{my_ref:luby_and_mitz}.
Their approach uses a peeling-style decoder based on verification
rather than erasures. Also, the algorithms in \cite{my_ref:luby_and_mitz}
are proposed in a node-based (NB) style but analyzed using message-based
(MB) decoders. It is implicitly assumed that the two approaches are
equivalent. In fact, this is not always true. In this paper, we consider
the differences between NB and MB decoders and derive an asymptotic
analysis for NB decoders.

The paper is organized as follows. In Section \ref{sec:descp_ana}, we describe the
LMP algorithm and use density evolution (DE) \cite{my_ref:DE} to analyze its performance. The difference
between NB and MB decoders for the first (LM1) and second algorithm
(LM2) in \cite{my_ref:luby_and_mitz} is discussed and the NB decoder
analysis is derived in Section \ref{sec:nb_ana}, respectively. The error floor of the LMP algorithms is considered in Section \ref{sec:err_flr}.
In Section \ref{sec:cmp_opt}, differential
evolution is used to optimize code ensembles and simulation results are
compared with the theoretical thresholds.
The results are also compared with previously published results from
\cite{my_ref:luby_and_mitz} and \cite{my_ref:isit2004wangshok}.
In Section \ref{sec:sim}, simulation results are presented. Applications of the
LMP algorithm are discussed and conclusions are given in Section \ref{sec:con}.

\section{Description and Analysis}
\label{sec:descp_ana}

\subsection{Description of the Decoding Algorithm}

\label{sec:decoder} The LMP decoder we discuss is designed mainly
for the $q$-SC and is based on local decoding operations applied
to lists of messages containing probable codeword symbols. The list messages passed in the
graph have three types: \emph{verified} (V), \emph{unverified} (U)
and \emph{erasure} (E). Every V-message has a symbol value associated
with it. Every U-message has a list of symbols associated with
it. Following \cite{my_ref:luby_and_mitz}, we mark messages
as verified when they are very likely to be correct. In particular,
we will find that the probability of FV approaches zero as $q$ goes
to infinity.

The LMP decoder works by passing list-messages around the decoding
graph. Instead of passing a single code symbol (e.g., Gallager A/B
algorithm \cite{my_ref:r1}) or a probability distribution over all
possible code symbols (e.g., \cite{my_ref:Davey}), we pass a list
of values that are more likely to be correct than the other messages.
At a check node, the output list contains all symbols which could
satisfy the check constraint for the given input lists. At the check
node, the output message will be verified if and only if all the incoming
messages are verified. At a node of degree $d$, the associativity
and commutativity of the node-processing operation allow it to be
decomposed into $(d-1)$ basic%
\footnote{Here we use {}``basic{}`` to emphasize that it maps two list-messages to a single list message.%
} operations (e.g., $a\!+\! b\!+\! c\!+\! d\!=\!(a\!+\! b)\!+\!(c\!+\! d)$).
In such a scheme, the computational complexity of each basic operation
is proportional to $s^{2}$ at the check node and $s\ln s$ at the
variable node%
\footnote{The basic operation at the variable node can be done by $s$ binary
searches of length $s$ and the complexity of a binary search of length
$s$ is $O(\ln s)$%
}, where $s$ is the list size of the input list. The list size grows
rapidly as the number of iterations increases. In order to make the
algorithm practical, we have to truncate the list to keep the list
size within some maximum value, denoted $S_{max}$. In our
analysis, we also find that, after the number of iterations exceeds
half the girth of the decoding graph, the probability of FV increases
very rapidly. We analyze the reasons of FV and classify the FV's into
two types. We find that the codes described in \cite{my_ref:luby_and_mitz}
and \cite{my_ref:isit2004wangshok} both suffer from type-II FV. In
Section \ref{sec:err_flr}, we analyze these FV's and propose a scheme that reduces the
probability of FV.


The message-passing decoding algorithm using list messages (or LMP)
applies the following simple rules to calculate the output messages
for a check node: 
\begin{itemize}
\item If all the input messages are verified, then the output becomes verified
with the value which makes all the incoming messages sum to zero. 
\item If any input message is an erasure, then the output message becomes
an erasure. 
\item If there is no erasure on the input lists, then the output list contains
all symbols which could satisfy the check constraint for the given
input lists. 
\item If the output list size is larger than $S_{max}$, then the output
message is an erasure. 
\end{itemize}
It applies the following rules to calculate the output messages of
a variable node: 
\begin{itemize}
\item If all the input messages are erasures or there are multiple verified
messages which disagree, then output message is the channel received
value. 
\item If any of the input messages are verified (and there is no disagreement)
or a symbol appears more than once, then the output message becomes
verified with the same value as the verified input message or the
symbol which appears more than once. 
\item If there is no verified message on the input lists and no symbol appears
more than once, then the output list is the union of all input lists. 
\item If the output message has list size larger than $S_{max}$, then the
output message is the received value from the channel. 
\end{itemize}
\vspace{-1mm}

\subsection{DE for Unbounded List Size Decoding Algorithm}

To apply DE to the LMP decoder with unbounded list sizes, denoted
LMP-$\infty$ (i.e., $S_{max}=\infty$), we consider three quantities
which evolve with the iteration number $i$. Let $x_{i}$ be the probability
that the correct   symbol is not on the list passed from a variable
node to a check node. Let $y_{i}$ be the probability that the message
passed from a variable node to a check node is not verified. Let $z_{i}$
be the average list size passed from a variable node to a check node.
The same variables are {}``marked'' $(\tilde{x}_{i},\tilde{y}_{i},\tilde{z}_{i})$
to represent the same values for messages passed from the check nodes
to the variable nodes (i.e., the half-iteration value). We also assume
all the messages are independent, that is, we assume that the bipartite graph
has girth greater than twice the number of decoding iterations.

First, we consider the probability, $x_{i}$, that the correct  
symbol is not on the list. For any degree-$d$ check node, the correct
message symbol will only be on the edge output list if all of the
other $d-1$ input lists contain their corresponding correct symbols.
This implies that $\tilde{x}_{i}=1-\rho(1-x_{i}$). For any degree-$d$
variable node, the correct message symbol is not on the edge
output list only if it is not on any of the other $d-1$ edge input lists.
This implies that $x_{i+1}=p\lambda(\tilde{x}_{i})$. This behavior
is very similar to erasure decoding of LDPC codes on the BEC and gives
the identical update equation \vspace{-1mm}
 \begin{equation}
\label{eq1}
x_{i+1}=p\lambda\left(1-\rho(1-x_{i})\right)
\end{equation}
 where $p$ is the $q$-SC error probability. Note that throughout the DE analysis, we assume that $q$ is sufficiently large. Next, we consider the
probability, $y_{i}$, that the message is not verified. For any degree-$d$
check node, an edge output message is verified only if all of the
other $d-1$ edge input messages are verified. For any degree-$d$
variable node, an edge output message is verified if any symbol on
the other $d-1$ edge input lists is verified or occurs twice which
implies $\tilde{y}_{i}=1-\rho(1-y_{i})$. The event that the output
message is not verified can be broken into the union of two disjoint
events: (i) the correct symbol is not on any of the input lists, and
(ii) the symbol from the channel is incorrect and the correct symbol
is on exactly one of the input lists and not verified. For a degree-$d$
variable node, this implies that \vspace{-1mm}
 \begin{equation}
\Pr(\textrm{not verified})=\left(\tilde{x}_{i}\right)^{d-1}+p(d-1)\left(\tilde{y}_{i}-\tilde{x}_{i}\right)\left(\tilde{x}_{i}\right)^{d-2}.\label{eq2}\end{equation}
 Summing over the d.d. gives the update equation 
\begin{multline}\label{eq3}
y_{i+1}= \lambda\left(1-\rho(1-x_{i})\right)+ \\ p\left(\rho(1-x_{i})-\rho(1-y_{i})\right)\lambda'\left(1-\rho(1-x_{i})\right).\end{multline}
 It is important to note that (\ref{eq1}) and (\ref{eq3}) were published
first in \cite[Thm. 2]{my_ref:isit2004wangshok} (by mapping $x_{i}=p_{i}$
and $y_{i}=p_{i}+q_{i}$), but were derived independently by us.

Finally, we consider the average list-size $z_{i}$. For any degree-$d$
check node, the output list size is equal%
\footnote{It is actually upper bounded because we ignore the possibility of
collisions between incorrect entries, but the probability of this
occurring is negligible as $q$ goes to infinity.%
}~to the product of the sizes of the other $d-1$ input lists. Since
the mean of the product of i.i.d. random variables is equal to the
product of the means, this implies that $\tilde{z}_{i}=\rho(z_{i})$.
For any degree-$d$ variable node, the output list size is equal to
one%
\footnote{A single symbol is always received from the channel.%
}~plus the sum of the sizes of the other $d-1$ input lists if the
output is not verified and one otherwise. Again, the mean of the sum
of $d-1$ i.i.d. random variables is simply $d-1$ times the mean
of the distribution, so the average output list size is given by 
\[
1+\left(\left(\tilde{x}_{i}\right)^{d-1}+p(d-1)\left(\tilde{y}_{i}-\tilde{x}_{i}\right)\left(\tilde{x}_{i}\right)^{d-2}\right)(d-1)\tilde{z}_{i}.\]
 This gives the update equation \begin{align*}
z_{i+1}= & 1\!+\!\left[\tilde{x}_{i}\lambda'\left(\tilde{x}_{i}\right)\!+\! p\left(\tilde{y}_{i}\!-\!\tilde{x}_{i}\right)\left(\lambda'\left(\tilde{x}_{i}\right)\!+\!\tilde{x}_{i}\lambda''\left(\tilde{x}_{i}\right)\right)\right]\rho(z_{i}).\end{align*}
 For the LMP decoding algorithm, the threshold of an ensemble $(\lambda(x),\rho(x))$
is defined to be \[
p^{*}\triangleq\sup\left\{ p\in(0,1]\bigg{|}p\lambda(1-\rho(1-x))<x\;\forall\; x\in(0,1]\right\} .\]
 Next, we show that some codes can achieve channel capacity using
this decoding algorithm. \begin{theorem} \label{thm1} Let $p^{*}$
be the threshold of the d.d. pair $(\lambda(x),\rho(x))$ and assume
that the channel error rate $p$ is less than $p^{*}$. In this case,
the probability $y_{i}$ that a message is not verified in the $i$-th
decoding iteration satisfies $\lim_{i\rightarrow\infty}y_{i}\rightarrow0$.
Moreover, for any $\epsilon>0$, there exists a $q<\infty$ such that
LMP decoding of a long random $(\lambda,\rho)$ LDPC code, on a $q$-SC
with error probability $p$, results in a symbol error rate less
than $\epsilon$. \end{theorem} \begin{IEEEproof} See Appendix~\ref{app_thm1}.
\end{IEEEproof} \begin{remark} Note that the convergence condition,
$p^{*}\lambda(1-\rho(1-x))<x$ for $x\in(0,1]$, is identical to the
BEC case but that $x$ has a different meaning. In the DE equation
for the $q$-SC, $x$ is the probability that the correct value is not
on the list. In the DE equation for the BEC, $x$ is the probability
that the message is an erasure. This tells us any capacity-achieving
ensemble for the BEC is capacity-achieving for the $q$-SC with LMP-$\infty$
algorithm and large enough $q$. This also gives some intuition about the
behavior of the $q$-SC for large $q$. For example, when $q$ is
large, an incorrectly received value behaves like an erasure \cite{my_ref:ShokITW04}.
\end{remark} 

\begin{corollary} \label{thm:lowrate} The code with d.d. pair $\lambda(x)=x$
and $\rho(x)=(1-\epsilon)x+\epsilon x^{2}$ has a threshold of $1-\frac{\epsilon}{1+\epsilon}$
and a rate of $r>\frac{\epsilon}{3(1+\epsilon)}$. Therefore, it achieves
a rate of $\Theta(\delta)$ for a channel error rate of $p=1-\delta$.
\end{corollary} \begin{IEEEproof} Follows from $\big(1-\frac{\epsilon}{1+\epsilon}\big)\lambda\left(1-\rho(1-x)\right)<x$
for $x\in(0,1]$ and Theorem \ref{thm1}. \end{IEEEproof}

\begin{remark} We believe that Corollary \ref{thm:lowrate} provides
the first linear-time decodable construction of rate $\Theta(\delta)$
for a random-error model with error probability $1-\delta$. A discussion
of linear-time encodable/decodable codes, for both random and adversarial
errors, can be found in \cite{my_ref:Guruswami03}. The complexity
also depends on the required list size which may be extremely large (though
independent of the block length). Unfortunately, we do not have explicit
bounds on the alphabet size or list size required for this construction.
\end{remark}

In practice, one cannot implement a list decoder with unbounded list
size. Therefore, we also evaluate the LMP decoder under a bounded
list-size assumption.

\subsection{DE for the Decoding Algorithm with Bounded List Size}

First, we introduce some definitions and notation for the DE analysis with
bounded list-size decoding algorithm. Note that, in the bounded list-size LMP algorithm, each list may contain at most $S_{max}$
symbols. For convenience, we classify the messages into four types: 
\begin{description}
\item [{(V)}] \emph{Verified}: message is verified and has list-size 1. 
\item [{(E)}] \emph{Erasure}: message is an erasure and has list-size 0. 
\item [{(L)}] \emph{Correct on list}: message is not verified or erased
and the correct symbol is on the list. 
\item [{(N)}] \emph{Correct not on list}: message is not verified or erased,
and the correct symbol is not on the list. 
\end{description}
For the first two message types, we only need to track the fractions,
$V_{i}$ and $E_{i}$, of message types in the $i$-th iteration.
For the third and the fourth types of messages, we also need to track
the list sizes. Therefore, we track the generating function of
the list size for these messages, given by $L_{i}(x)$ and $N_{i}(x)$.
The coefficient of $x^{j}$ represents the probability that the message
has list-size $j$. Specifically, $L_{i}(x)$ is defined by \[
L_{i}(x)=\sum_{j=1}^{S_{max}}l_{i,j}x^{j},\]
 where $l_{i,j}$ is the probability that, in the $i$-th decoding
iteration, the correct symbol is on the list and the message list
has size $j$. The function $N_{i}(x)$ is defined similarly. This
implies that $L_{i}(1)$ is the probability that the list contains
the correct symbol and that it is not verified. For the same reason,
$N_{i}(1)$ gives the probability that the list does not contain the
correct symbol and that it is not verified. For the simplicity of expression,
we denote the overall density as $P_{i}=[V_{i},E_{i},L_{i}(x),N_{i}(x)]$. The same variables are
{}``marked'' $(\tilde{V},\tilde{E},\tilde{L},\tilde{N}$ and $\tilde{P})$
to represent the same values for messages passed from the check nodes
to the variable nodes (i.e., the half-iteration value). 

Using these definitions, we find that DE can be computed efficiently
using polynomial arithmetic. For convenience of analysis
and implementation, we use a sequence of basic operations plus a
separate truncation operator to represent a multiple-input multiple-output
operation. We use $\boxplus$ to denote the check-node operator and
$\otimes$ to denote the variable-node operator. Using this, the DE
for the variable-node basic operation $P^{(3)}=\tilde{P}^{(1)}\otimes\tilde{P}^{(2)}$
is given by \vspace{-2mm}


\begin{align}
V^{(3)}= & \, \tilde{V}^{(1)}+\tilde{V}^{(2)}-\tilde{V}^{(1)}\tilde{V}^{(2)}+\tilde{L}^{(1)}(1)\tilde{L}^{(2)}(1)\label{eq8} \\
E^{(3)}= & \, \tilde{E}^{(1)}\tilde{E}^{(2)}\label{eq9} \\
L^{(3)}(x)  = & \, \tilde{L}^{(1)}(x)\left(\tilde{E}^{(2)} + \tilde{N}^{(2)}(x)\right) \nonumber \\
&+\tilde{L}^{(2)}(x)\left(\tilde{E}^{(1)}+\tilde{N}^{(1)}(x)\right) \label{eq10} \\
N^{(3)}(x)= & \, \tilde{N}^{(1)}(x)\tilde{E}^{(2)}+\tilde{N}^{(2)}(x)\tilde{E}^{(1)} \nonumber \\
&+\tilde{N}^{(1)}(x)\tilde{N}^{(2)}(x).\label{eq11}
\end{align}
 Note that  (\ref{eq8})-(\ref{eq11}) do not yet consider
the list-size truncation and the channel value.
 For the basic check-node operation $\tilde{P}^{(3)}=P^{(1)}\boxplus P^{(2)}$,
the DE is given by
\begin{align}
\tilde{V}^{(3)}= & \, V^{(1)}V^{(2)}\\
\tilde{E}^{(3)}= & \, E^{(1)}+ E^{(2)}-E^{(1)}E^{(2)} \\
\tilde{L}^{(3)}(z)= & \, \left[V^{(1)}L^{(2)}(z)+ V^{(2)}L^{(1)}(z) \right. \nonumber \\
& \, \left. + \, L^{(1)}(x)L^{(2)}(y)\right]_{x^{j}y^{k}\rightarrow z^{jk}} \\
\tilde{N}^{(3)}(z)= & \, \left[N^{(1)}(x)N^{(2)}(y)+ N^{(1)}(x)\left(V^{(2)}y+ L^{(2)}(y)\right)\right.\nonumber \\
 & \, \left. + \, N^{(2)}(x)\left(V^{(1)}y+ L^{(1)}(y)\right)\right]_{x^{j}y^{k}\rightarrow z^{jk}}
\end{align}
 where the subscript $x^{j}y^{k}\rightarrow z^{jk}$ denotes the substitution
of variables. Finally, the truncation of lists to size $S_{max}$
is handled by truncation operators which map densities to densities.
We use $\mathcal{T}$ and $\mathcal{T'}$ to denote the truncation
operation at the check and variable nodes. Specifically, we truncate
terms with degree higher than $S_{max}$ in the polynomials $L(x)$
and $N(x)$. At check nodes, the truncated probability mass is moved
to $E$.

\noindent At variable nodes, lists longer than $S_{max}$ entries are replaced
by the channel value. Let  $P'_i=\big( \tilde{P}_{i}^{\otimes k-1} \big)$ be the intermediate density which is the 
result of applying the basic operation $k-1$ times on $\tilde{P}_{i}$.
After considering the channel value and list truncation, the symbol node message density is given by $\mathcal{T'}(P'_i)$. To analyze this, we separate $L'_i(x)$ into two
terms: $ {A'_i}(x)$ with degree less than $S_{max}$ and $x^{S_{max}}{B'_i}(x)$
with degree at least $S_{max}$. Likewise, we separate ${N'_i}(x)$
into $ {C'_i}(x)$ and $x^{S_{max}} {D'_i}(x)$. The inclusion
of the channel symbol and the truncation are combined into a single
operation \[
{\textstyle P_i\!=\!\mathcal{T}'\!\left(\left[ {V'_i}, {E'_i}, {A'_i}(x)\!+\!x^{S_{max}}{B'_i}(x), {C'_i}(x)\!+\!x^{S_{max}} {D'_i}(x)\right]\right)}\]
 defined by \begin{align}
V_i= & \, {V'_i}\!+\!(1-p)\left( {A'_i}(1)+ {B'_i}(1)\right)\label{eq17}\\
E_i= & \,0\\
L_i(x)= & \,(1-p)x\left({E'_i}\!+\! {C'_i}(x)\!+\! {D'_i}(1)\right)\!+\! px {A'_i}(x)\\
N_i(x)= & \, px\left( {E'_i}\!+\! {B'_i}(1)\!+\! {C'_i}(x)\!+\! {D'_i}(1)\right).\end{align}
 Note that in (\ref{eq17}), the term $(1-p)\left( {A'_i}(1)+ {B'_i}(1)\right)$
is due to the fact that messages are compared for possible verification
before truncation.

The overall DE recursion is easily written in terms of the forward
(symbol to check) density $P_{i}$ and the backward (check to symbol) density
$\tilde{P}_{i}$ by taking the irregularity into account. The initial density is $P_{0}=[0,0,(1-p)x,px]$,
where $p$ is the error probability of the $q$-SC channel, and the
recursion is given by 
\begin{align}
\tilde{P}_{i}= & \sum_{k=2}^{d_{c}}\rho_{k}\,\mathcal{T}\left(P_{i}^{\boxplus k-1}\right)\\
P_{i+1}= & \sum_{k=2}^{d_{v}}\lambda_{k}\,\mathcal{T'}\left(\tilde{P}_{i}^{\otimes k-1}\right).\end{align}
 Note that the DE recursion is not one-dimensional. This makes it
difficult to optimize the ensemble analytically. It remains an open problem
to find the closed-form expression of the threshold in terms of the
maximum list size, d.d. pairs, and the alphabet size $q$. In section \ref{sec:cmp_opt}, we will fix
the maximum variable and check degrees, code rate, $q$ and maximum
list size and optimize the threshold over the d.d. pairs by using
a numerical approach.

\begin{section}{Analysis of Node-based Algorithms}
\label{sec:nb_ana}
\begin{subsection}{Differential Equation Analysis of LM1-NB}

We refer to the first and second algorithms in \cite{my_ref:luby_and_mitz}
as LM1 and LM2, respectively. Each algorithm can be viewed either
as message-based (MB) or node-based (NB). The first and second algorithms
in \cite{my_ref:isit2004wangshok} and \cite{my_ref:Shokwangpersonal}
are referred to as SW1 and SW2. These algorithms are summarized in
Table~\ref{tab:table3}. Note that, if no verification occurs,
the variable node (VN) sends the ({}``channel value'', U) and the check node (CN) sends the ({}``expected
correct value'',U) in all these algorithms. The algorithms SW1, SW2 and LMP are all MB algorithms,
but can be modified to be NB algorithms.

\subsubsection{Motivation}

\begin{small} %
\begin{table}[t]

\caption{Brief Description of Message-Passing Algorithms for $q$-SC}

\label{tab:table3} \centering \begin{tabular}{|c|l|}
\hline 
\textbf{Alg.}  & \hspace{1in} \textbf{Description} \tabularnewline
\hline 
\multirow{1}{*}{LMP-$S_{max}$}  & LMP as described in Section~\ref{sec:decoder} with maximum \\ & list-size $S_{max}$ \tabularnewline
\hline 
\multirow{1}{*}{LM1-MB}  & MP decoder that passes (value, $U$/$V$). \cite[III.B]{my_ref:luby_and_mitz} \tabularnewline
 & At VN's, output is $V$ if any input is $V$ or message  \\ & matches  channel value, otherwise pass channel value.\tabularnewline
 & At CN's, output is $V$ iff all inputs are $V$.  \tabularnewline
\hline 
\multirow{1}{*}{LM1-NB}  & Peeling decoder with VN state (value, $U$/$V$). \cite[III.B]{my_ref:luby_and_mitz} \tabularnewline
 & At CN's, if all neighbors sum to 0, all neighbors get $V$. \tabularnewline
 & At CN's, if all neighbors but one are $V$, then last is $V$. \tabularnewline
\hline 
\multirow{1}{*}{LM2-MB}  & The same as LM1-MB with one extra rule. \cite[IV.A]{my_ref:luby_and_mitz}. \tabularnewline
 & At VN's, if two input messages match, then output $V$. \tabularnewline
\hline 
\multirow{1}{*}{LM2-NB}  & The same as LM1-NB with one extra rule. \cite[IV.A]{my_ref:luby_and_mitz}. \tabularnewline
 & At VN's, if two neighbor values same, then VN gets $V$. \tabularnewline
\hline 
\multirow{1}{*}{SW1}  & Identical to LM2-MB\tabularnewline
\hline 
\multirow{1}{*}{SW2}  & Identical to LMP-$\infty$. \cite[Thm. 2]{my_ref:isit2004wangshok} \tabularnewline
\hline
\end{tabular}
\end{table}

\end{small}

In \cite{my_ref:luby_and_mitz}, the algorithms are proposed in the
node-based (NB) style \cite[Section III-A and IV]{my_ref:luby_and_mitz},
but analyzed in the message-based (MB) style \cite[Section III-B and IV]{my_ref:luby_and_mitz}.
It is easy to verify that the LM1-NB and LM1-MB have identical performance, but this is not true for the
NB and MB LM2 algorithms. In this section, we will
show the differences between the NB decoder and MB decoder and derive
a precise analysis for LM1-NB.

First, we discuss the equivalence of LM1-MB and LM1-NB.
\begin{theorem}\label{thm2}
Let $A$ be the set of variable nodes that are verified when LM1-NB decoding terminates.
For the same received sequence, let $B$ be the set of variable nodes that have at least one verified output message when LM1-MB terminates.
Then, LM1-NB and LM1-MB are equivalent in the sense that $A=B$.
\end{theorem}
\begin{IEEEproof}[Sketch of Proof]
Though the basic idea behind this result is relatively straightforward, the details are somewhat lengthy and, therefore, deferred to a more general treatment~\cite{Zhang-unpub11}.
The first observation is that the LM1-NB and LM1-MB decoders both satisfy a monotonicity property that, assuming no FV, guarantees convergence to a fixed point.
In particular, the messages are ordered with respect to type (e.g., verified $>$ correct $>$ incorrect) and vectors of messages are endowed with the induced partial ordering.
Using this approach, it can be shown that the fixed point messages of the LM1-NB decoder cannot be worse than those of the LM1-MB decoder.
Finally, a detailed analysis of the computation tree, for an arbitrary LM1-MB fixed point, shows that every node verified by LM1-NB must have at least one verified output message.
\end{IEEEproof}

In the NB decoder, the verification status is associated with a node.
Once a node is verified, all outgoing messages are verified.
In the MB decoder, the status is associated with an edge
and the outgoing message on each edge may have a different verification status.
NB algorithms cannot, in general, be analyzed using DE because the
independence assumption between messages does not hold. Therefore,
we develop peeling-style decoders, which are equivalent to LM1-NB and LM2-NB,
and use differential equations to analyze them.

It is worth noting that the threshold of a node-based algorithm is at least as large as its message-based counterpart.
This is because node-based processing can only result in additional verifications during each step.
The intuition behind this can be seen by looking at a variable node of degree 3.
The message-based decoder may sometimes output a verified message on only one edge.
This occurs, for example, if the channel value is incorrect and the three input messages are $V,I,I$.
But, the node-based decoder verifies all output edges in this case.
The drawback is that the correlation caused by node-based processing complicates the analysis because density evolution is based on the assumption that all input messages are independent.


Following \cite{my_ref:lubyca1}, we analyze the peeling-style decoder
using differential equations that track the average number of edges
(grouped into types) in the graph as decoding progresses. From the
results of \cite{my_ref:lubyca1} and \cite{my_ref:diffeqn}, we
know that the actual number of edges (of any type), in any particular
decoding realization is tightly concentrated around the average over
the lifetime of the random process. For peeling-style decoding,
a variable node and its edges are removed after verification;
each check node keeps track of its new parity constraint (i.e., the
value to which the attached variables must sum) by subtracting values
associated with the removed edges.

\subsubsection{Analysis of Peeling-Style Decoding}

First, we introduce some notation and definitions for the analysis.
A variable node (VN) whose channel value is correctly received is
called a correct variable node (CVN), otherwise it is called an incorrect
variable node (IVN). A check node (CN) with $i$ edges connected to
the CVN's and $j$ edges connected to the IVN's will be said to have
C-degree $i$ and I-degree $j$, or type $n_{i,j}$. We note that the analysis of node-based algorithms 
hold for irregular LDPC code ensemble of maximum variable node degree $d_v$ and maximum check node degree $d_c$.

We also define the following quantities: 
\begin{itemize}
\item $t$: decoding time or fraction of VNs removed from graph 
\item $L_{i}(t)$: the number of edges connected to CVN's with degree $i$
at time $t$
\item $R_{j}(t)$: the number of edges connected to IVN's with degree $j$
at time $t$
\item $N_{i,j}(t)$: the number of edges connected to CN's with C-degree
$i$ and I-degree $j$
\item $E_{l}(t)$: the remaining number of edges connected to CVN's at time
$t$
\item $E_{r}(t)$: the remaining number of edges connected to IVN's at time
$t$
\item $a(t)$: the average degree of CVN's which have at least 1 edge coming from CN's of type $n_{i,1},i\ge 1$, \[
a(t)=\frac{\sum_{k=1}^{d_v}{kL_k(t)}}{E_l(t)}\]
 
\item $b(t)$: the average degree of IVN's which have at least 1 edge coming from CN's of type $n_{0,1}$, \[
b(t)=\frac{\sum_{k=1}^{d_v}{kR_k(t)}}{E_r(t)}\]
 
\item $E$: number of edges in the original graph, \[
E=E_{l}(0)+E_{r}(0).\]
 
\end{itemize}
Counting edges in three ways gives the identity
\[ \sum_{i\ge1}L_{i}(t)+\sum_{i\ge1}R_{i}(t)=E_{l}(t)+E_{r}(t)= \!\!\!\! \sum_{\substack{i\geq0,j\geq0\\(i,j)\neq(0,0)}} \!\!\!\! N_{i,j}(t).\]

These r.v.'s represent a particular realization of the decoder. The
differential equations are defined for the normalized (i.e., divided
by $E$) expected values of these variables. We use lower-case notation
(e.g., $l_{i}(t)$, $r_{i}(t)$, $n_{i,j}(t)$\footnote{When we use $n_{i,j}$, we refer to the type of IVN's. When we use $n_{i,j}(t)$, we refer to the normalized expected values.}, etc.) for these deterministic
trajectories. Time is scaled so that the decoder removes exactly one
variable node during $1/N$ time units, where $N$ is the block length of the code.

The description of peeling-style decoder is as follows. The peeling-style
decoder removes one CVN or IVN in each time step by the following
rules: 
\begin{description}
\setlength{\labelwidth}{11mm}
\setlength{\itemindent}{2mm}
\item [ CER:]If any CN has all its edges connected to CVN's, remove
one of these CVN's and all of its edges. 
\item [ IER1:]If any IVN has at least one edge connected to a
CN of type $(0,1)$, then compute the value of the IVN from the attached
CN and remove the IVN along with all its edges. 
\end{description}
If both CER and IER1 can be applied, then one is chosen randomly as
described below.

\begin{figure}[t]
\centering
\includegraphics[width=0.35\columnwidth,angle=270,viewport=150 180 450 600]{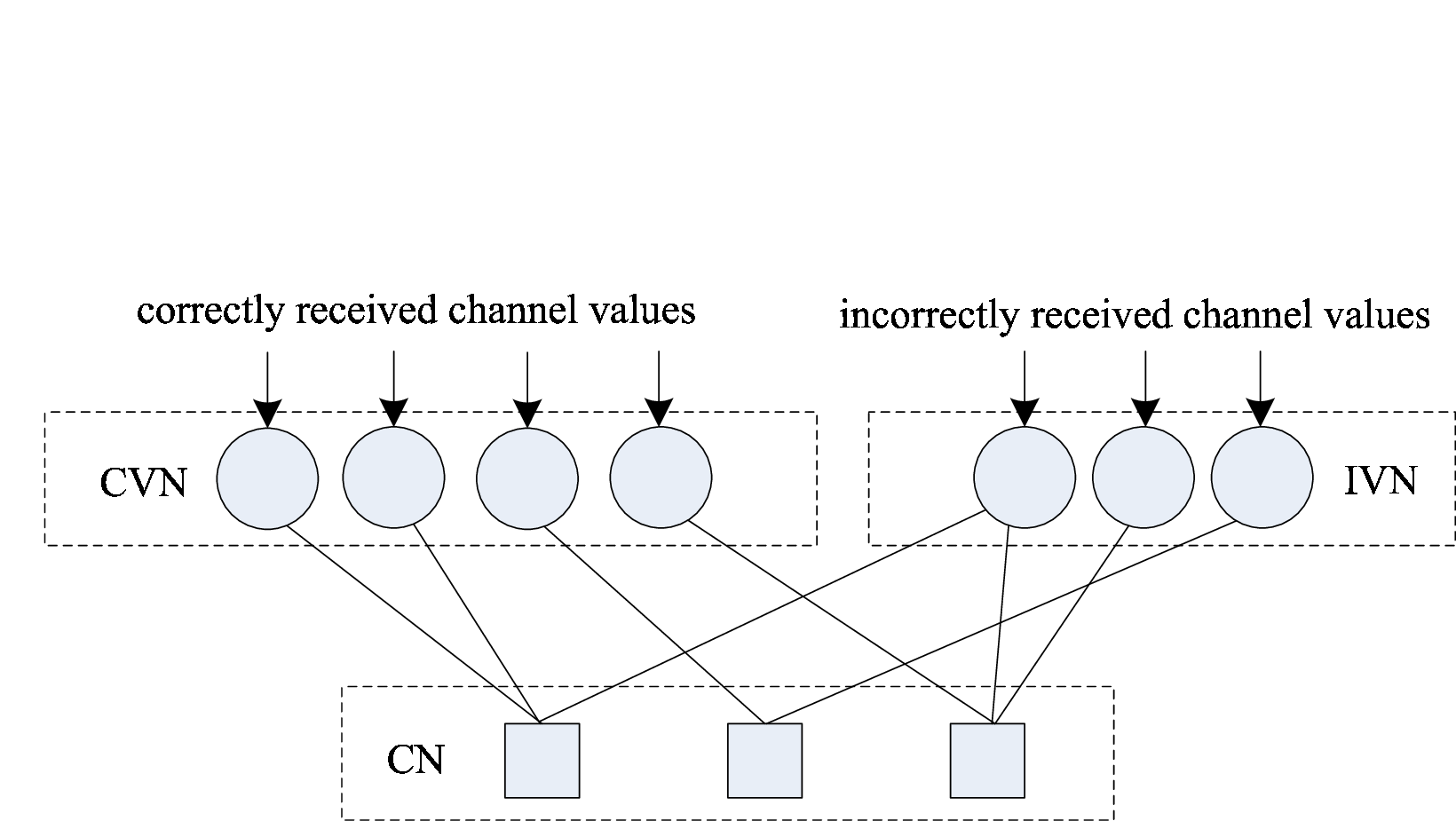}
\caption{Tanner graph for the LM1 differential equation analysis.}\label{fig:fig3}
\end{figure}

Since both rules remove exactly one VN, the decoding process either
finishes in exactly $N$ steps or stops early and cannot continue.
The first case occurs only when either the IER1 or CER condition is
satisfied in every time step. When the decoder stops early, the pattern
of CVNs and IVNs remaining is called a stopping set. We also note
that the rules above, though described differently, are equivalent
to the first node-based algorithm (LM1-NB) introduced in \cite{my_ref:luby_and_mitz}.

Recall that the node-based algorithm for LM1, from \cite{my_ref:luby_and_mitz},
has two verification rules.
The first rule is: if the neighbors (of a CN) have values that sum to zero,
then all the neighbors are verified to their current values.
In this analysis, however, the neighbors are verified one at a time.
Since this implies that the CN is attached only to CVNs (assuming no FV),
we call this correct-edge-removal (CER) and notice that it is only allowed
if $n_{i,0}>0$ for some $i\ge1$.
The second rule is: if all neighbors (of a CN) are verified except for one,
then the last neighbor is verified to the value that satisfies the check.
If the last neighbor is an IVN, we call this type-I incorrect-edge-removal
(IER1) and notice that it is only allowed when $n_{0,1}(t)>0$. 

The peeling-style decoder performs one operation during each time step.
The operation is random and can be either CER or IER1. When both operations are possible,
we choose randomly between these two rules by picking CER with probability
$c_{1}(t)$ and IER1 with probability $c_{2}(t)$, where \begin{align*}
c_{1}(t) & =\frac{\sum_{i\geq1}n_{i,0}(t)}{\sum_{i\geq1}n_{i,0}(t)+n_{0,1}(t)}\\
c_{2}(t) & =\frac{n_{0,1}(t)}{\sum_{i\geq1}n_{i,0}(t)+n_{0,1}(t)}.\end{align*}
This weighted sum ensures that the expected change in the decoder state is Lipschitz continuous if either $c_{1}(t)$ or $c_{2}(t)$ is strictly positive.
Therefore, the differential equations can be written as \begin{align*}
\frac{\mbox{d}l_{i}(t)}{\mbox{d}t} & =c_{1}(t)\frac{\mbox{d}l_{i}^{(1)}(t)}{\mbox{d}t}+c_{2}(t)\frac{\mbox{d}l_{i}^{(2)}(t)}{\mbox{d}t}\\
\frac{\mbox{d}r_{i}(t)}{\mbox{d}t} & =c_{1}(t)\frac{\mbox{d}r_{i}^{(1)}(t)}{\mbox{d}t}+c_{2}(t)\frac{\mbox{d}r_{i}^{(2)}(t)}{\mbox{d}t}\\
\frac{\mbox{d}n_{i,j}(t)}{\mbox{d}t} & =c_{1}(t)\frac{\mbox{d}n_{i,j}^{(1)}(t)}{\mbox{d}t}+c_{2}(t)\frac{\mbox{d}n_{i,j}^{(2)}(t)}{\mbox{d}t},\end{align*}
 where $^{(1)}$ and $^{(2)}$ denote, respectively, the effects of
CER and IER1.

\subsubsection{CER Analysis}

If the CER operation is picked, then we choose randomly an edge attached
to a CN of type $(i,0)$ with $i\geq1$. This VN endpoint of this
edge is distributed uniformly across the CVN edge sockets. Therefore,
it will be attached to a CVN of degree $k$ with probability ${l_{k}(t)}/{e_{l}(t)}$.
Therefore, one has the following differential equations for $l_{k}$
and $r_{k}$ \[
\frac{\mbox{d}l_{k}^{(1)}(t)}{\mbox{d}t}=\frac{l_{k}(t)}{e_{l}(t)}(-k),\textrm{ for }k\ge1\]
 and \[
\frac{\mbox{d}r_{k}^{(1)}(t)}{\mbox{d}t}=0.\]

For the effect on check edges, we can think of removing a CVN with
degree $k$ as first randomly picking an edge of type $(k,0)$ connected
to that CVN and then removing all the other $k-1$ edges (called reflected
edges) attached to the same CVN. The $k-1$ reflected edges are uniformly
distributed over the $E_{l}(t)$ correct sockets of the CN's.
These $k-1$ reflected edges hit $\frac{n_{i,j}(t)i(k-1)}{(i+j)e_{l}(t)}$ CN's of type $(i,j)$ on average.
Next, we average over the VN degree, $k$, and find that
\[ p_{i,j}^{(1)}(t) \triangleq \frac{n_{i,j}(t)i(a(t)-1)}{(i+j)e_{l}(t)}\]
is the expected number of type-$(i,j)$ CN's hit by reflected edges.

If a CN of type $(i,j)$ is hit by a reflected edge, then the decoder state loses $i+j$
edges of type $(i,j)$ and gains $i-1+j$ edges of type $(i-1,j)$.
Hence, one has the following differential equation, for $j>0$ and
$i+j\le d_{c}$, \[
\frac{\mbox{d}n_{i,j}^{(1)}(t)}{\mbox{d}t}=\left(p_{i+1,j}^{(1)}(t)-p_{i,j}^{(1)}(t)\right)(i+j).\]
One should keep in mind that $n_{i,j}(t)=0$ for $i+j>d_{c}$.

For $n_{i,j}^{(1)}(t)$ with $j=0$, the effect from above must be
combined with effect of the type-$(i,0)$ initial edge that was chosen.
So the differential equation becomes \[
\frac{\mbox{d}n_{i,0}^{(1)}(t)}{\mbox{d}t}=\left(p_{i+1,0}^{(1)}(t)-p_{i,0}^{(1)}(t)\right)i+\left(q_{i+1}^{(1)}(t)-q_{i}^{(1)}(t)\right)i\]
 where \[
q_{i}^{(1)}(t) \triangleq \frac{n_{i,0}(t)}{\sum_{m\geq1}n_{m,0}(t)}.\] Note that $p^{(1)}_{d_c+1,0}(t)\triangleq0$ and $q^{(1)}_{d_c+1}(t	)\triangleq0$

\subsubsection{IER1 Analysis}

If the IER1 operation is picked, then we choose a random CN of type
$(0,1)$ and follow its only edge to the set of IVNs. The edge is attached
uniformly to this set, so the differential equations for IER1 are given by
\[ \frac{\mbox{d}l_{k}^{(2)}(t)}{\mbox{d}t}=0,\]
 \[
\frac{\mbox{d}r_{k}^{(2)}(t)}{\mbox{d}t}=\frac{r_{k}(t)}{e_{r}(t)}(-k),\]
 and
\[ \frac{\mbox{d}n_{i,j}^{(2)}(t)}{\mbox{d}t}=\left(p_{i,j+1}^{(2)}(t)-p_{i,j}^{(2)}(t)\right)(i+j) - \delta_{i,0} \delta_{j,1}, \]
where \[
p_{i,j}^{(2)}(t) \triangleq \frac{n_{i,j}(t)j(b(t)-1)}{(i+j)e_{r}(t)}.\]
We note that the removal of the initial edge, when $(i,j)=(0,1)$, is taken into account by the Kronecker delta functions.


Notice that even for (3,6) codes, there are 30 differential equations%
\footnote{There are 28 for $n_{i,j}(t)$ ($i,j\in[0,\cdots,6]$ such that $i+j\le6$),
1 for $r_{k}(t)$, and 1 for $l_{k}(t)$.%
} to solve. So we solve the differential equations numerically and
the threshold for (3,6) code with LM1 is $p^{*}=0.169$. This coincides
with the result from density evolution analysis for LM1-MB in \cite{my_ref:luby_and_mitz}
and hints at the equivalence between LM1-NB and LM1-MB. In the proof
of Theorem \ref{thm2} we make this equivalence precise by showing
that the stopping sets of LM1-NB and LM1-MB are the same. \end{subsection}

\begin{subsection}{Differential Equation Analysis of LM2-NB}
Similar to the analysis of  LM1-NB algorithm, we analyze  LM2-NB algorithm
by analyzing a peeling-style decoder which is equivalent to the LM2-NB
decoding algorithm. The peeling-style decoder removes one CVN or IVN
during each time step according to the following rules:
\begin{description}
\item [{CER:}] If any CN has all its edges connected to CVN's, pick one
of the CVN's and remove it.
\item [{IER1:}] If any IVN has any messages from CN's with type $n_{0,1}$,
then the IVN and all its outgoing edges can be removed and we track
the correct value by subtracting the value from the check node.
\item [{IER2:}] If any IVN is attached to more than one CN with I-degree
1, then it will be verified and all its outgoing edges can be removed.
 
\end{description}
For simplicity, we first introduce some definitions and short-hand notation:
\begin{itemize}
\item Correct edges: edges which are connected to CVN's
\item Incorrect edges: edges which are connected to IVN's
\item CER edges: edges which are connected to check nodes with type
$n_{i,0}$ for $i\ge1$ 
\item IER1 edges: edges which are connected to check nodes with type
$n_{0,1}$
\item IER2 edges: edges which connect IVN's and the check nodes with
type $n_{i,1}$ for $i\ge1$
\item NI edges: normal incorrect edges, which are incorrect edges but neither
IER1 edges nor IER2 edges
\item CER nodes: CVN's which have at least one CER edge
\item IER1 nodes: IVN's which have at least one IER1 edge
\item IER2 nodes: IVN's which have at least two IER2 edges
\item NI nodes: IVN's which contain at most 1 IER2 edge and no IER1 edges.
\end{itemize}
Note that an IVN can be both an IER1 node and an IER2 node at the
same time. 

The analysis of LM2-NB is much more complicated than LM1-NB because
the IER2 operation makes the distribution of IER2 edges dependent
on each other. For example, the IER2 operation removes a randomly chosen
IVN with more than 2 IER2 edges; this reduces the number of IVNs which
have multiple IER2 edges and therefore the remaining IER2 edges are more
likely to land on different IVN's. 

The basic idea behind our analysis of the LM2-NB decoder is that we can
separate the incorrect edges into types and assume that mapping between sockets is given by
a uniform random permutation.  Strictly speaking, this is not true and another
approach, which leads to the same differential equations, is used when
considering a formal proof of correctness.
 In detail, we model the
structure of an LDPC code during LM2-NB decoding as shown in Fig. \ref{fig:v(d)}
with one type for correct edges and three types for incorrect edges.
The following calculations assume the four permutations, labeled CER, NI, IER2, and
IER1, are all uniform random permutations.

\begin{figure}
\vspace{2mm}
\begin{center}
\includegraphics[width=0.9\columnwidth]{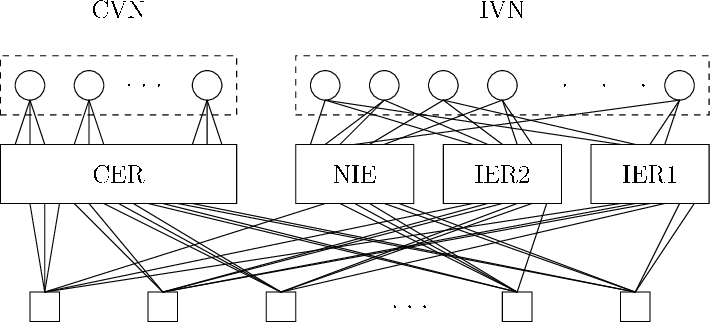}
\end{center}
\vspace{-2mm}
\caption{\label{fig:v(d)}Graph structure of an LDPC code during LM2-NB decoding.}
\end{figure}

The peeling-style decoder randomly chooses one VN from the set of
CER IER1 and IER2 nodes and removes this node and all its edges at
each step. The idea of the analysis is to first calculate the probability
of choosing a VN with a certain type, i.e., CER, IER1 or IER2, and
the node degree. We then analyze how removing this VN affects the
system parameters. 

In the analysis, we will track the evolution of the following system parameters:
\begin{itemize}
\item $l_{k}(t)$: the fraction of edges connected to CVN's with degree
$k$, $0<k\le d_{v}$ at time $t$
\item $r_{i,j,k}(t):$ the fraction of edges connected to IVN's with $i$
NI edges, $j$ IER2 edges and $k$ IER1 edges at time $t$, $i,j,k\in\{0,1,\dots,d_{v}\}$
and $0<i+j+k\le d_{v}$
\item $n_{i,j}(t):$ the fraction of edges connected to check nodes with
$i$ correct edges and $j$ incorrect edges at time $t$, $i,j\in\{0,1,\dots,d_{c}\}$
and $0<i+j\le d_{c}$.
\end{itemize}
We note that, when we say ``fraction'', we mean the number of
a certain type of edges/nodes normalized by the number of edges/nodes
in the \emph{original graph}. 

The following quantities can be calculated from $l_{k}(t)$, $r_{i,j,k}(t)$ and $n_{i,j}(t)$:
\begin{itemize}
\item $e_{l}(t)\triangleq\sum_{k=1}^{d_{v}}l_{k}(t)$: the fraction of correct
edges
\item $e_{r}(t)\triangleq\sum_{i=0}^{d_{v}}\sum_{j=0}^{d_{v}-i}\sum_{k=0}^{d_{v}-i-j}r_{i,j,k}(t)$:
the fraction of incorrect edges
\item $\eta_{0}(t)\triangleq\sum_{j=2}^{d_{c}}\sum_{i=0}^{d_{c}-j}\frac{jn_{i,j}(t)}{i+j} \\ =\sum_{i=1}^{d_{v}}\sum_{j=0}^{d_{v}-i}\sum_{k=0}^{d_{v}-i-j}\frac{ir_{i,j,k}(t)}{i+j+k}$:
the fraction of NI edges
\item $\eta_{1}(t)\triangleq n_{0,1}(t)=\sum_{k=1}^{d_{v}}\sum_{j=0}^{d_{v}-k}\sum_{i=0}^{d_{v}-i-j}\frac{kr_{i,j,k}(t)}{(i+j+k)}$:
the fraction of IER1 edges
\item $\eta_{2}(t)\triangleq\sum_{i=1}^{d_{c}}\frac{n_{i,1}(t)}{(i+1)} \\ =\sum_{j=1}^{d_{v}}\sum_{i=0}^{d_{v}-j}\sum_{k=0}^{d_{v}-i-j}\frac{jr_{i,j,k}(t)}{(i+j+k)}$:
the fraction of IER2 edges
\item $s_{0}(t)\triangleq\sum_{j=0}^{1}\sum_{i=1}^{d_{v}-j}\frac{r_{i,j,0}}{i+j}$:
the fraction of NI nodes
\item $s_{1}(t)\triangleq\sum_{k=1}^{d_{v}}\frac{n_{k,0}(t)}{k}$: the fraction
of CER nodes
\item $s_{2}(t)\triangleq\sum_{k=1}^{d_{v}}\sum_{i=0}^{d_{v}-k}\sum_{j=0}^{d_{v}-i-k}\frac{r_{i,j,k}}{i+j+k}$:
the fraction of IER1 nodes
\item $s_{3}(t)\triangleq\sum_{j=2}^{d_{v}}\sum_{i=0}^{d_{v}-j}\sum_{k=0}^{d_{v}-i-j}\frac{r_{i,j,k}}{i+j+k}$:
the fraction of IER2 nodes.
\end{itemize}
As in the LM1-NB analysis, we use superscript $^{(1)}$ to denote
the contribution of the CER operations. We use $^{(2)}$ to denote
the contribution of the IER1 operations and $^{(3)}$ to denote the
contribution of the IER2 operations. Since we assume that the decoder
randomly chooses a VN from the set of CER, IER1 and IER2 nodes and
removes all its edges during each time step, the differential equations
of the system parameters can be written as the weighted sum of the
contributions of the CER, IER1 and IER2 operations. The weights are chosen
to be \begin{align*}
c_{1}(t) & =\frac{s_{1}(t)}{(s_{1}(t)+s_{2}(t)+s_{3}(t))}\\
c_{2}(t) & =\frac{s_{2}(t)}{(s_{1}(t)+s_{2}(t)+s_{3}(t))}\\
c_{3}(t) & =\frac{s_{3}(t)}{(s_{1}(t)+s_{2}(t)+s_{3}(t))}.\end{align*}
This weighted sum ensures that the expected change in the decoder state is Lipschitz continuous if any one $c_{1}(t)$, $c_{2}(t)$, or $c_{3} (t)$ is strictly positive.
 Next, we will show how CER, IER1 and IER2 operations affect the system
parameters. 

Given the d.d. pair $(\lambda,\rho)$ and the channel error probability
$p$, we initialize the state as follows. Since a fraction $(1-p)\lambda_{k}$
of the edges are connected to CVN's of degree $k$, we initialize
$l_{k}(t)$ with \[
l_{k}(0)=(1-p)\lambda_{k},\]
for $k=1,2,\dots d_{v}$. Noticing that each CN socket is connected
to a correct edge with probability $(1-p)$ and incorrect edge with
probability $p$, we initialize $n_{i,j}(t)$ with\[
n_{i,j}(0)=\rho_{i+j}\binom{i+j}{i}(1-p)^{i}p^{j},\]
for $i+j\in\{1,2,\dots,d_{c}\}$.  The probability that an IVN socket
is connected to an NI, IER1 edge, or IER2 edge is denoted respectively
by $g_0$, $g_1$, or $g_2$ with
\begin{align*}
g_0 &= \frac{1}{p}\sum_{j'=2}^{d_{c}}\sum_{i'=0}^{d_{c}-j'}\frac{j'n_{i',j'}(0)}{i'+j'} \\
g_1 &= \frac{1}{p} \sum_{i'=1}^{d_{c}}\frac{n_{i',1}(0)}{(i'+1)} \\
g_2 &= \frac{1}{p} n_{0,1}(0).
\end{align*}
Therefore, we initialize $r_{i,j,k}(t)$ with\[
r_{i,j,k}(0)=p\lambda_{i+j+k}\binom{i+j+k}{i,j,k} g_{0}^i g_{1}^j g_{2}^k ,\]
for $i+j+k\in\{1,2,\dots,d_{v}\}$.

\subsubsection{CER analysis}

The analysis for $\frac{\textrm{d}l_{k}^{(1)}(t)}{\textrm{d}t}$ is
the same as the LM1-NB analysis. In the CER operation, the decoder randomly
selects a CER edge.
A CVN with degree $k$ is chosen with probability ${l_{k}(t)}/{e_{l}(t)}$.
If a degree $k$ CVN is chosen, the number of edges of type $l_{k}$ decreases by $k$ and therefore
\[ \frac{\mbox{d}l_{k}^{(1)}(t)}{\mbox{d}t}=\frac{-kl_{k}(t)}{e_{l}(t)}.\]
For $j\ge1$ and $i+j\le d_{c}$ \[
\frac{\mbox{d}n_{i,j}^{(1)}(t)}{\mbox{d}t}=\left(p_{i+1,j}^{(1)}(t)-p_{i,j}^{(1)}(t)\right)(i+j)\]
where $a(t)=\frac{\sum_{k=1}^{d_{v}}kl_{k}(t)}{e_{l}(t)}$ is the
average degree of the CVN's which are hit by the initially chosen
CER edge and $p_{i,j}^{(1)}=\frac{n_{i,j}(t)i(a(t)-1)}{(i+j)e_{l}(t)}$
is the average number of CN's with type $n_{i,j}$ hit by the $a(t)-1$
reflecting edges. 

For $j=0$ and $i\ge1$, we also have to consider the initially chosen
CER edge. This gives \[
\frac{\mbox{d}n_{i,0}^{(1)}(t)}{\mbox{d}t}=\left(p_{i+1,0}^{(1)}(t)-p_{i,0}^{(1)}(t)\right)i+\left(q_{i+1}^{(1)}(t)-q_{i}^{(1)}(t)\right)i\]
 where $q_{i}^{(1)}(t)=\frac{n_{i,0}(t)}{\sum_{m\ge1}n_{m,0}(t)}$
is the probability that the initially chosen CER edge is of type $n_{i,0}$. 

When the removed CER node has a reflecting edge that hits a CN
of type $n_{1,1}$, the CN's IER2 edge becomes an IER1 edge. This is the
only way the CER operation can affect $r_{i,j,k}.$ On average, each
CER operation generates $a(t)-1$ reflecting edges. For each reflecting
edge, the probability that it hits a CN of type $n_{1,1}$ is $\frac{n_{1,1}(t)}{2e_{l}(t)}$.
Taking this IER2 to IER1 conversion into account, for $k \geq 1$ and $j<d_v$, gives
\begin{align*}
&\frac{\textrm{d} r_{i,j,k}^{(1)}(t)}{\mbox{d}t}
= \, (a(t)\!-\!1) \frac{n_{1,1}(t)}{2e_{l}(t)}\left(\frac{jr_{i,j,k}(t)}{(i\!+\!j\!+\!k)\eta_{2}(t)}(-(i\!+\!j\!+\!k))\right. \\ 
& \quad \quad \quad \quad \quad \left. -\frac{(j+1)r_{i,j+1,k-1}(t)}{(i\!+\!j\!+\!k)\eta_{2}(t)}(-(i\!+\!j\!+\!k))\right)\\
& \quad = (a(t)\!-\!1) \frac{n_{1,1}(t)}{2e_{l}(t)}\left(\frac{-jr_{i,j,k}(t)}{\eta_{2}(t)}+\frac{(j\!+\!1)r_{i,j+1,k-1}(t)}{\eta_{2}(t)}\right).
\end{align*}
If $k=0$ or $j=d_v$, then the IVN's with type $r_{i,j,k}$
can only lose edges and \[
\frac{\textrm{d}r_{i,j,k}^{(1)}(t)}{\textrm{d}t}=(a(t)-1)\frac{n_{1,1}(t)}{2e_{l}(t)}\left(\frac{-jr_{i,j,k}(t)}{\eta_{2}(t)}\right).\]

\subsubsection{IER2 analysis}

Since the IER2 operation does not affect $l_{k}(t)$, we have \[
\frac{\mbox{d}l_{k}^{(3)}(t)}{\mbox{d}t}=0.\]
To analyze how the IER2 operation changes $n_{i,j}(t)$ and $r_{i,j,k}(t)$,
we first observe that the probability a randomly chosen IER2 node
is of type $r_{i,j,k}$ is given by \[
\Pr\left(\mbox{type }r_{i,j,k}\,|\,\mbox{IER2 node}\right)=\frac{\frac{r_{i,j,k}(t)}{i+j+k}}{s_{3}(t)},\]
 for $j\ge2$. Otherwise, $\Pr\left(\mbox{type }r_{i,j,k}\,|\,\mbox{IER2 node}\right)=0$. 

Let us denote the contribution to $\frac{\textrm{d}n_{i',j'}^{(3)}(t)}{\textrm{d}t}$ caused by removing:
one NI edge as $u_{i',j'}(t)$,
one IER2 edge as $v_{i',j'}(t)$,
and one IER1 edge as $w_{i',j'}(t)$. Then, we can
write the total contribution to the derivative as
\begin{multline*}
\frac{\textrm{d}n_{i',j'}^{(3)}(t)}{\textrm{d}t} =\sum_{i=0}^{d_{v}}\sum_{j=0}^{d_{v}}\sum_{k=0}^{d_{v}}\Pr\left(\mbox{type }r_{i,j,k}\,|\,\mbox{IER2 node}\right) \\ \left(iu_{i',j'}(t)+jv_{i',j'}(t)+kw_{i',j'}(t)\right).\end{multline*}

First, we consider $u_{i',j'}(t)$. If an NI edge is chosen from
the IVN side, then it hits a CN of type $n_{i,j}$ with probability $\frac{jn_{i,j}(t)}{\eta_{0}(i+j)}$
if $j\ge2$ and probability 0 otherwise. When $2 \leq j\le d_{v}-1$,
we have \begin{align*}
u_{i',j'}(t) & =\frac{j'n_{i',j'}(t)(-(i'+j'))}{(i'+j')\eta_{0}(t)}+\frac{(j'+1)n_{i',j'+1}(i'+j')}{(i'+j'+1)\eta_{0}(t)}\\
 & =\frac{-j'n_{i',j'}(t)}{\eta_{0}(t)}+\frac{(j'+1)n_{i',j'+1}(i'+j')}{(i'+j'+1)\eta_{0}(t)}.\end{align*}
When $j=d_{v}$, one finds instead that \[
u_{i',j'}(t)=\frac{-j'n_{i'j'}(t)}{\eta_{0}(t)}.\]

Since an NI edge cannot be connected to a CN of type $n_{i',1}$,
we must treat $j'=1$ separately. Notice that type $n_{i',1}$ may still
gain edges from type $n_{i',2}$, so we have\[
u_{i',1}(t)=\frac{2n_{i',2}(t)(i'+1)}{(i'+2)\eta_{0}(t)}.\]
When $j'=0$, CN's with type $n_{i',0}$ do not have any NI edges, so we have  $u_{i',0}(t)=0$.
Now we consider $v_{i',j'}(t)$. Since edges of type $n_{i',j'}$
with $j\ge2$ cannot be IER2 edges, $n_{i'j'}(t)$ with $j\ge2$ is not
affected by removing IER2 edges. The IER2 edge removal reduces the
number of edges of type $n_{i',1}$, $i\ge1$, so we have $v_{i',1}(t)=-\frac{n_{i',1}(t)}{\eta_{2}(t)}$.
When $j'=0$ and $i'\ge1$, we have $v_{i',0}(t)=\frac{n_{i',1}(t)}{\eta_{2}(t)}$.
Only CN's with type $n_{0,1}$ are affected when we remove an
IER1 edge on the IVN side. So we have $w_{0,1}(t)=1$
and $w_{i',j'}(t)=0$ when $(i',j')(t)\neq(0,1).$ 

Next, we derive the contribution to the $r_{i,j,k}(t)$ differential equation caused
by removing an IER2 node. When the decoder removes an IER2 node of
type $r_{i',j',k'}$, there are direct and indirect effects.
The direct effect is the loss of $i'$ NI edges, $j'$ IER2 edges and $k'$ IER1 edges.
This edge removal affects CN degrees and
the type of some CN edges may also change and thereby indirectly affect $r_{i,j,k}(t)$.
We call these indirectly affected edges ``CN reflected edges''.

Let $u'_{i,j,k}(t)$ be the contribution of CN reflected edges to $\frac{\textrm{d}r_{i,j,k}(t)}{\textrm{d}t}$
caused by the removal of: an NI edge be $u'_{i,j,k}(t)$, an IER2 edge be $v'_{i,j,k}(t)$, and an IER2 edge be $w'_{i,j,k}(t)$.
Then, we can write the total contribution to the derivative as
\begin{multline*}
\frac{\textrm{d}r_{i,j,k}^{(3)}(t)}{\textrm{d}t}  =-\Pr\left(\mbox{type }r_{i,j,k}\,|\,\mbox{IER2 node}\right)(i+j+k)\,+\\
  \sum_{i'=0}^{d_{v}}\sum_{j'=0}^{d_{v}}\sum_{k'=0}^{d_{v}}\Pr\left(\mbox{type }r_{i',j',k'}\,|\,\mbox{IER2 node}\right) \\ \left(i'u'_{i,j,k}(t)+j'v'_{i,j,k}(t)+k'w'_{i,j,k}(t)\right).\end{multline*}

There are two ways that the CN reflected edges of an NI edge can
affect $r_{i,j,k}(t)$. The first occurs when the CN is of type $n_{i,2}$ and $1\le i\le d_{c}-2$.
In this case, removing an NI edge changes the type of the other
incorrect edge from NI to IER2. The second occurs when the CN is
of type $n_{0,2}$. Removing an NI edge changes the type of the other
incorrect edge from NI to IER1.
The probability that an NI edge hits a CN of type $n_{i,2},$ (for $1\le i\le d_{c}-2$) is $\frac{\sum_{i=1}^{d_{c}-2}\frac{2n_{i,2}(t)}{i+2}}{\eta_{0}(t)}$.
Likewise, the probability that an NI edge hits a CN of type $n_{0,2}$ is $\frac{n_{0,2}(t)}{\eta_{0}(t)}$.
Since the probability that an NI edge is connected to an IVN of type $r_{i,j,k}$ is $\frac{ir_{i,j,k}(t)}{(i+j+k)\eta_{0}(t)}$, one finds that
\begin{align*}
u'_{i,j,k}(t) & \! = \! \frac{\sum_{i=1}^{d_{c}-2}\frac{2n_{i,2}(t)}{i+2}}{\eta_{0}(t)}\! \left(\!-\frac{ir_{i,j,k}(t)}{\eta_{0}(t)}\!+\!\frac{(i\!+\!1)r_{i+1,j-1,k}(t)}{\eta_{0}(t)}\right)\\
 & +\frac{n_{0,2}(t)}{\eta_{0}(t)}\left(-\frac{ir_{i,j,k}(t)}{\eta_{0}(t)}+\frac{(i+1)r_{i+1,j,k-1}(t)}{\eta_{0}(t)}\right),
\end{align*}
where $r_{i,j,k}(t)\triangleq 0$ unless $i,j,k \in \{ 0,\ldots,d_v \}$ and $i+j+k\leq d_v$.
Since there are no CN reflected edges of type IER1 and IER2, one also finds that $v'_{i,j,k}(t)=0$
and $w'_{i,j,k}(t)=0$.

\subsubsection{IER1 Analysis}

Like the IER2 operation, the IER1 operation does not affect $l_{k}(t)$ and therefore
\[ \frac{\textrm{d}l_{k}^{(2)}(t)}{\textrm{d}t}=0.\]
To analyze how IER1 changes $n_{i,j}(t)$ and $r_{i,j,k}(t)$, we
observe that the probability a randomly chosen IER1 node is
of type $r_{i,j,k}$ is given by
\[ \Pr\left(\mbox{type }r_{i,j,k}\,|\,\mbox{IER1 node}\right)=\frac{\frac{r_{i,j,k}(t)}{i+j+k}}{s_{2}(t)}\]
when $k\ge1$.  Otherwise, $\Pr\left(\mbox{type }r_{i,j,k}\,|\,\mbox{IER1 node}\right)=0$.

Applying the same arguments used for the IER2 operation, one finds that, \begin{multline*}
\frac{\textrm{d}n_{i',j'}^{(2)}(t)}{\textrm{d}t}=\sum_{i=0}^{d_{v}}\sum_{j=0}^{d_{v}}\sum_{k=0}^{d_{v}}\Pr\left(\mbox{type }r_{i,j,k}\,|\,\mbox{IER1 node}\right) \\ \left(iu_{i',j'}(t)+jv_{i',j'}(t)+kw_{i',j'}(t)\right)\end{multline*}
 and \begin{multline*}
\frac{\textrm{d}r_{i,j,k}^{(2)}(t)}{\textrm{d}t}  =-\Pr\left(\mbox{type }r_{i',j',k'}\,|\,\mbox{IER1 node}\right)(i'+j'+k')
 \\ + \sum_{i'=0}^{d_{v}}\sum_{j'=0}^{d_{v}}\sum_{k'=0}^{d_{v}}\Pr\left(\mbox{type }r_{i',j',k'}\,|\,\mbox{IER1 node}\right) \\ \left(i'u'_{i,j,k}(t)+j'v'_{i,j,k}(t)+k'w'_{i,j,k}(t)\right).\end{multline*}
The program used to perform these computations and compute LM2-NB thresholds is available online \cite{Zhang-web09}.

\subsection{Discussion of the Analysis}

Consider the peeling decoder for the BEC introduced in \cite{my_ref:luby_and_mitz}.
Throughout the decoding process, one reveals and then removes edges
one at a time from a hidden random graph. The analysis of this decoder
is simplified by the fact that, given the current residual degree
distribution, the unrevealed portion of the graph remains uniform
for every decoding trajectory. In fact, one can build a finite-length
decoding simulation never constructs the actual decoding graph. Instead,
it tracks only the residual degree distribution of the graph and implicitly
chooses a random decoding graph one edge at a time.

For asymptotically long codes,\cite{my_ref:luby_and_mitz} used this approach 
to derive an analysis based on differential equations. This analysis
is actually quite general and can also be applied to other peeling-style
decoders in which the unrevealed graph is not uniform. One may observe
this from its proof of correctness, which depends only on two important
observations. First, the distribution of all decoding paths is concentrated
very tightly around its average as the system size increases. Second,
the expected change in the decoder state can be written as a Lipschitz
function of the current decoder state. If one augments the decoding
state to include enough information so that the expected change can
be computed from the augmented state (even for non-uniform residual
graphs), then the theorem still applies. 

The differential equation computes the average evolution, over all
random bipartite graphs, of the system parameters as
the block length $n$ goes to infinity. While the numerical simulation
of long codes gives the evolution of the system parameters of a particular
code (a particular bipartite graph) as $n$ goes to infinity. To prove
that the differential equation analysis precisely predicts the evolution
of the system parameters of a particular code, one must show the
concentration of the evolution of the system parameters of a particular
code around the ensemble average as $n$ goes to infinity. 

In the LM2-NB algorithm, one node is removed at a time but this can also
be viewed as removing each edge sequentially. The main difference
for LM2-NB algorithm is that one has more edge types and one must track
some details of the edge types on both the check nodes and the variable
nodes. This causes a significant problem in the analysis because updating
the exact effect of edge removal requires revealing some edges before
they will be removed. For example, the CER operation can cause an
IER2 edge to become an IER1 edge, but revealing the adjacent symbol
node (or type) renders the analysis intractable. 

Unfortunately, our proof of correctness still relies on two unproven
assumptions which we state as a conjectures.  This section leverages the
framework of \cite{my_ref:luby_and_mitz,my_ref:diffeqn,wormald_de_1997}
by describing only the new discrete-time random process $H_t$ associated
with our analysis.

We first introduce the definitions of the random process. In this subsection, 
we use $t$ to represent the discrete time.  We
follow the same notation used in \cite{my_ref:luby_and_mitz}. Let the life span of
the random process be $\alpha_0 n$. Let $\Omega$ denote a probability space
and $S$ be a measurable space of observations. A discrete-time random
process over $\Omega$ with observations $S$ is a sequence $Q\triangleq(Q_{0},Q_{1},\dots)$
of random variables where $Q_{t}$ contains the information revealed
at $t$-th step. We denote the history of the process up to time $t$
as $H_{t}\triangleq(Q_{0},Q_{1},\dots,Q_{t}).$ Let $S^{+}:=\cup_{i\ge1}S^{i}$
denote the set of all histories and $\mathcal{Y}$ be the set of all
decoder states. One typically uses a state space that tracks the number
of edges of a certain type (e.g., the degree of the attached nodes). 

We define the random process as follows.  The total number of edges connected to 
IVN's with type $r_{i,j,k}$ at time $t$ is denoted $R_{i,j,k}(t)$ and the total number of edges connected to 
check nodes with type $n_{i,j}$ is $N_{i,j}(t)$.
The main difference is that we track the average $\bar{R}_{i,j,k}(t)\triangleq E[R_{i,j,k}(t)|H_t]$ of node degree distribution rather than the exact value.

Let $R(t)$, $\bar{R}(t)$, and $N(t)$ be vectors of random variables formed by including all valid $i,j,k$ tuples for each variable.
Using this, the decoder state at time $t$ is given by $Y_{t}\triangleq\{N(t),\bar{R}(t)\}$.
To connect this with \cite[Theorem~5.1]{wormald_de_1997}, we define the
history of our random process as follows. In the beginning of the
decoding, we label the variable/check nodes by their degrees. When the
decoder removes an edge, the revealed information $Q_{t}$ contains
the degree of the variable node and type of the check node to which the removed
edge is connected. We note that sometimes the edge-removal operation changes the
type of the unremoved edge on that check node. In this case, $Q_{t}$
also contains the information about the type of the check node to which this CN reflected edge connects.
But $Q_{t}$ does not contain any information about the
IVN to which the CN-reflecting edge is connected. By defining the
history in this manner, $Y_t$ is a deterministic function of $H_{t}$
and can be made to satisfy the conditions of \cite[Theorem~5.1]{wormald_de_1997}.

The following conjecture, which basically says that $R_{i,j,k}(t)$ concentrates, encapsulates
one of the unproven assumptions needed to establish the correctness of this analysis.
\begin{conject}
\[ \lim_{n\rightarrow \infty}\Pr\left(\sup_{0\le t\le \alpha_0 n}\left|\bar{R}_{i,j,k}(t)-R_{i,j,k}(t)\right|\ge n^{5/6}\right)=0 \]
holds for all $i,j,k \in \{0,\ldots,d_v\}$ such that $i+j+k \leq d_v$.
\end{conject}

The next observation is that the expected drift $E[Y_{t+1}-Y_t|H_t]$ can be
computed exactly in terms of $R(t)$ if the four edge-type permutations
are uniform.  But, only $\bar{R}(t)$ can be computed exactly from $H_t$.
Let $f(Y_t)$ denote the expected drift under the uniform assumption using
$\bar{R}(t)$ instead of $R(t)$.  Since $R(t)$ is concentrated around
$\bar{R}(t)$, by assumption, and $f$ is Lipschitz, this is not the main
difficulty.  Instead, the uniform assumption is problematic and
the following conjecture sidesteps the problem by assuming that the
true expected drift $E[Y_{t+1}-Y_t|H_t]$ is asymptotically equal
to $f(Y_t)$.
\begin{conject}
\[\lim_{n\rightarrow \infty}\!\! \Pr\left( \sup_{0\le t\le \alpha_0 n}\lVert E[Y_{t+1}\!-\!Y_t|H_t] \!-\! f(Y_t) \rVert_{\infty} \ge n^{-1/7}\right)\!=\!0.\]
\end{conject}


If these conjectures hold true, then \cite[Theorem~5.1]{wormald_de_1997}
can be used to show that the differential equation correctly models the
expected decoding trajectory and actual realizations concentrate tightly
around this expectation.  In particular, we find that $N_{i,j}(nt)$ concentrates
around $n_{i,j}(t)$ and both $\bar R_{i,j,k}(nt)$ and $R_{i,j,k}(nt)$ concentrate
around $r_{i,j,k}(t)$.  These conclusions are supported by our simulations.

\end{subsection}
\end{section}

\section{Error Floor Analysis of LMP Algorithms}
\label{sec:err_flr}
During the simulation of the optimized ensembles from Table~\ref{tab:opt_our},
we observed an unexpected error floor. While one might expect an error floor due to finite $q$ effects,
it was somewhat surprising that the error floor persisted even when $q$ was chosen to be large enough
so that no FV's were observed in the error floor regime.
Analyzing the simulation results shows that the error floor was caused instead by symbols that remain
unverified at the end of decoding.
This motivated further study into the error floor of LMP algorithms. It is also worth noting that,
when $q$ is relatively small, the error floor is caused by several factors including as type-I FV, type-II FV (which 
we will discuss later) and the event that some symbols remain unverified when the decoding terminates. 
For small $q$, these factors appear to interact in a complex manner.

In this section, we focus only on error floors due to unverified symbols.
There are three reasons for this.
The first reason is that this was the dominant event we saw in our simulations (i.e., the error floors
observed in the simulation are not caused by FV). The second reason is that
the assumption that ``verified symbols are correct with high probability" is the cornerstone of verification-based decoding
and very little can be said about verification decoding without this assumption.
For example, if FV has any significant impact, then both the density-evolution analysis and the differential-equation 
analysis break down and the thresholds become meaningless.
The last reason is for simplicity; one can analyze the error floors
caused by each factor separately in this case because they are not strongly coupled. 

Although the dominant contribution to the error floor is not 
caused by FV, an analysis of FV is provided for sake of the completeness. This analysis actually 
helps us  understand why the dominant error events caused by FV can be avoided by increasing $q$.

\subsection{The Union Bound for ML Decoding}

First, we derive the union bound on the probability of error with ML decoding
for the $q$-SC. To match our simulations with the union bounds, we
also expurgate our ensemble to remove all codeword weights that have an expected
multiplicity less than 1.

Next, we summarize a few results from \cite[p.~497]{RU-2008}
that characterize the low-weight codewords of LDPC codes with degree-2
variable nodes. When the block length is large, all of these low-weight
codewords are caused, with high probability, by short cycles of degree-2
nodes. For binary codes, the number of codewords with weight $k$
is a random variable which converges to a Poisson distribution with
mean $\frac{1}{2k}(\lambda_{2}\rho^{'}(1))^{k}$. When the
channel quality is high (i.e., high SNR, low error/erasure rate),
the probability of ML decoding error is mainly caused by low-weight
codewords.

For non-binary $GF(q)$ codes, a codeword is supported on a cycle
of degree-2 nodes only if the product of the edge weights is 1. This
occurs with probability $1/(q-1)$ if we choose the i.i.d. uniform
random edge weights for the code. Hence, the number of $GF(q)$ codewords
of weight $k$ is a random variable, denoted $B_{k}$, which converges
to a Poisson distribution with mean $b_{k}=\frac{1}{2k(q-1)}(\lambda_{2}\rho^{'}(1))^{k}$.
After expurgating the ensemble to remove all weights with expected multiplicity less
than 1, $k_{1}=\min \{ k\geq1 | b_{k}^{(n)}\ge1\}$ becomes the minimum
codeword weight. An upper bound on the pairwise error probability (PEP) of the $q$-SC with error probability
$p$ is given by the following lemma. 
\begin{lemma}
\label{lemma3} Let $y$ be the received symbol sequence assuming the
all-zero codeword was transmitted.  Let $u$ be any codeword with exactly
$k$ non-zero symbols. Then, the probability that the ML decoder chooses $u$ over
the all-zero codeword is upper bounded by
\[p_{2,k} \leq \left(p\frac{q-2}{q-1} +\sqrt{\frac{4p(1-p)}{q-1}} \right)^k.\]
 
\end{lemma}
\begin{IEEEproof} See Appendix~\ref{app_lemma3}. \end{IEEEproof}

\begin{remark} Notice that $b_{k}$ is exponential in $k$ and the
PEP is also exponential in $k$. The union bound for the frame error
rate, due to low-weight codewords, can be written as \[
P_{B}\le\sum_{k=k_{1}}^{\infty}b_{k}p_{2,k}.\]
 It is easy to see $k_{1}=\Omega(\log q)$ and the sum is dominated by
the first term $b_{k_{1}}p_{2,k_{1}}$ which has the smallest exponent.
When $q$ is large, the PEP upper bound is on the order of $O\left(p^{k}\right)$.
Therefore, the order of the union bound on frame error rate with ML
decoding is \[
P_{B}=O\left(\frac{\left(\lambda_{2}\rho^{'}(1)p\right)^{\log q}}{q\log q}\right)\]
 and the expected number of symbols in error is \[
O\left(\frac{\left(\lambda_{2}\rho^{'}(1)p\right)^{\log q}}{q}\right),\] if $p\lambda_2\rho'(1)<1.$
 \end{remark}

\subsection{Error Analysis for LMP Algorithms}

\label{false_v}

The error of LMP algorithm comes from two types of decoding failure.
The first type of decoding failure is due to unverified symbols. The
second one is caused by FV. To understand
the performance of LMP algorithms, we analyze these two types
separately. When we analyze each error type,
we neglect the interaction for simplicity.

FV's can be classified into two types. The first type is, as \cite{my_ref:luby_and_mitz}
mentions, when the error magnitudes in a single check sum to zero;
we call this type-I FV. For single-element lists, it occurs with probability
roughly $1/q$ (i.e., the chance that two uniform random symbols are
equal). For multiple lists with multiple entries, we analyze the FV
probability under the assumption that no list contains the correct
symbol. In this case, each list is uniform on the $q-1$ incorrect
symbols. For $m$ lists of size $s_{1},\ldots,s_{m}$, the type-I FV
probability is given by $1-\binom{q-1}{s_{1},s_{2},\cdots,s_{m}}\big/\prod_{i=1}^{m}\binom{q-1}{s_{i}}$.
In general, the Birthday paradox applies and the FV probability is
roughly $s^{2}\binom{m}{2}/q$ for large $q$ and equal size lists.

The second type of FV is that messages become more and more correlated
as the number of iterations grows, so that an incorrect message may
go through different paths and return to the same node. We denote
this kind of FV as a type-II FV.

\begin{figure}[t]
\centering

\includegraphics[width=0.6\columnwidth]{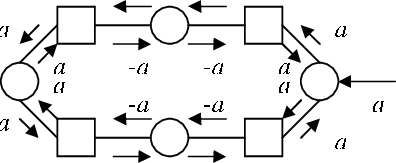}
\caption{An example of type-II FV's.}\label{fig2}
\end{figure}

These two types of FV are quite different.
One cannot avoid type-II FV by increasing $q$ without randomizing the edge weights and one cannot
avoid type-I FV by constraining the number of decoding iterations
to be within half of the girth (or increasing the girth). Fig.~\ref{fig2}
shows an example of type-II FV. In Fig.~\ref{fig2}, there is an
8-cycle in the graph (involving 4 variable nodes) and we assume the variable node on the right
has an incorrect incoming message {}``$a$''. Assume that the all-zero
codeword is transmitted, all the incoming messages at each variable
node are not verified, the list size is less than $S_{max}$, and
each incoming message at each check node contains the correct symbol.
In this case, the incorrect symbol will travel along the cycle and
cause FV's at all variable nodes along the cycle. If the characteristic
of the field is 2, there are a total of $c/2$ FV's occurring along
the cycle, where $c$ is the length of the cycle. This type of FV
can be reduced significantly by choosing each non-zero entry in the
parity-check matrix randomly from the non-zero elements of Galois
field. In this case, a cycle causes a type-II FV only if the product
of the edge-weights along that cycle is 1. Therefore, we suggest choosing
the non-zero entries of the parity-check matrix randomly to mitigate
type-II FV. Recall that the idea of using non-binary elements in the
parity-check matrix is quite old and appears in early work on LDPC codes over
$GF(q)$ \cite{my_ref:Davey}.

\subsection{An Upper Bound on Type-II FV Probability for Cycles}

In this subsection, we analyze the probability of error caused by
type-II FV. Note that type-II FV occurs only when the depth-$2k$
directed neighborhood of an edge (or node) has cycles. But type-I
FV occurs at every edge (or node). The order of the probability that
type-I FV occurs is approximately $O(1/q)$ \cite{my_ref:luby_and_mitz}.
The probability of type-II FV is hard to analyze because it depends
on $q$, $S_{max}$ and $k$ in a complicated way. But an upper bound
of the probability of the type-II FV is derived in this section.

Since the probability of type-II FV is dominated by short cycles of
degree-2 nodes, we only analyze type-II FV along these short cycles.
As we will soon see, the probability of type-II FV is exponential
in the length of the cycle. So, the error caused by type-II FV on
cycles is dominated by short cycles. We also assume $S_{max}$ to
be large enough such that an incorrectly received value can pass around
a cycle without being truncated. This assumption makes our analysis
an upper bound. Another condition required for an incorrectly received
value to participate in a type-II FV is that the product of the edge
weights along the cycle is 1. If we assume that almost all edges not
on the cycle are verified, then once any edge on the cycle is verified,
all edges will be verified in the next $k$ iterations. So we also
assume that nodes along a cycle are either all verified or all unverified.

We note that there are three possible patterns of verification on
a cycle, depending on the received values. The first case is that
all the nodes are received incorrectly. As mentioned above, the incorrect
value passes around the cycle without being truncated, comes back to 
the node again and falsely verifies the outgoing messages of the node.
So all messages will be falsely verified (if they are all received
incorrectly) after $k$ iterations. Note that this happens with probability
$\frac{1}{q-1}p^{k}.$ The second case is that all messages are verified
correctly, say, no FV. Note that this does not require
all the nodes to have correctly received values. For example, if any
pair of adjacent nodes are received correctly, it is easy to see all
messages will be correctly verified. The last case is that there is at
least 1 incorrectly received node in any pair of adjacent nodes and
that there is at least 1 node with correctly received value on the cycle.
In this case, all messages will be verified after $k$ iterations,
i.e., messages from correct nodes are verified correctly and those
from incorrect nodes are falsely verified. Then the verified messages
will propagate and half of the messages will be verified correctly
and the other half will be falsely verified. Note that this happens
with probability $\frac{1}{q-1}2\left(p^{k/2}-p^{k}\right)\approx\frac{2p^{k/2}}{q-1}$
and this approximation gives an upper bound even if we combine the
previous $\frac{1}{q-1}p^{k}$ term.

Recall that the number of cycles with $k$ variable nodes converges to a Poisson
with mean $\frac{1}{2k}(\lambda_{2}\rho'(1))^{k}$. Using
the union bound, we can upper bound on the ensemble average probability
of any type-II FV event with 
\begin{align*}
\Pr(\mbox{any}&\mbox{ type-II FV})
\le\sum_{k=k_{1}}^{\infty}\frac{\left(\lambda_{2}\rho^{'}(1)\right)^{k}}{2k(q-1)}2p^{\frac{k}{2}} \\
&=\sum_{k=k_{1}}^{\infty}\frac{\left(\lambda_{2}\rho^{'}(1)\sqrt{p}\right)^{k}}{k(q-1)}
=O\left(\frac{\left(\lambda_{2}\rho^{'}(1)\sqrt{p}\right)^{\log q}}{(q-1)\log q}\right).
\end{align*}
The ensemble average number of nodes involved in type-II FV events
is given by
\begin{align*}
\mbox{E}[\mbox{symbols}&\mbox{ in type-II FV}]
\le\sum_{k=k_{1}}^{\infty}\frac{\left(\lambda_{2}\rho^{'}(1)\right)^{k}}{2k(q-1)}2kp^{\frac{k}{2}} \\
&=\sum_{k=k_{1}}^{\infty}\frac{\left(\lambda_{2}\rho^{'}(1)\sqrt{p}\right)^{k}}{(q-1)}
= O\left(\frac{\left(\lambda_{2}\rho^{'}(1)\sqrt{p}\right)}{(q-1)}\right).
\end{align*}
Notice that both these upper bounds are decreasing functions of $q$.

\subsection{Upper Bound on Unverification Probability for Cycles}

During the simulation of the optimized ensembles from Table~\ref{tab:opt_our},
we observed significant error floors and all of the error events were caused by unverified symbols when the decoding terminates. 
In this subsection, We derive the union bound for the probability
of decoder failure caused by the symbols on short cycles which never
become verified. We call this event as \emph{unverification} and we denote it by UV. As described
above, to match the settings of the simulation and simplify the analysis, we assume $q$ is large enough to have arbitrarily small
probability of both type-I and type-II FV. In this case, the error
is dominated by the unverified messages because the following analysis
shows that the union bound on the probability of unverification is
independent of $q$.

In contrast to type-II FV, the unverification event does not require cycles,
i.e., unverification occurs even on subgraphs without cycles. But
in the low error-rate regime, the dominant contribution to unverification events
comes from short cycles of degree-2 nodes. Therefore, we only analyze the
probability of unverification caused by short cycles of degree-2 nodes.

Consider a degree-2 cycle with $k$ variable nodes and assume that no FV occurs
in the neighborhood of this cycle. Assuming the maximum list size
is $S_{max}$, the condition for UV
is that there is at most one correctly received value along $S_{max}+1$
adjacent variable nodes. Note that we don't consider type-II FV since
type-II FV occurs with probability $\frac{1}{q-1}$ and we can choose
$q$ to be arbitrarily large. On the other hand, unverification does
not require the product of the edge weights on a cycle to be 1, so
one cannot mitigate it by increasing $q$.
Let the r.v. $U$ be the number of symbols involved in unverification events on short cycles of degree-2 nodes.
One can use the union bound to upper bound the probability of any unverification events by
\[ \Pr (U\geq 1)\le\sum_{k= k_{2}}^{\infty}\frac{\left(\lambda_{2}\rho^{'}(1)\right)^{k}}{2k}\phi(S_{max},p,k)\]
 where $k_{2}=\min \{k\geq1 | \frac{1}{2k}(\lambda_{2}\rho^{'}(1))^{k}\ge1 \}$
and $\phi(S_{max},p,k)$ is the UV probability for a cycle consisting of $k$ degree-2 nodes.
One can also upper bound the the expectation of $U$ with
\begin{equation}
\mbox{E}[U] \le \sum_{k= k_{2}}^{\infty} \frac{\left(\lambda_{2}\rho^{'}(1)\right)^{k}}{2}\phi(S_{max},p,k).\label{ub_unver}
\end{equation}
The unverification probability for short cycles of degree-2 nodes is give by the following lemma. 
\begin{lemma}
\label{lemma53} Let the cycle have $k$ degree-2 variable nodes, the maximum
list size be $s$, and the channel error probability be $p$. Then, the probability
of an unverification event is $\phi(s,p,k)=\textrm{Tr}\left(B^{k}(p)\right)$ where
$B(p)$ is the $(s+1)$ by $(s+1)$ matrix 
\begin{equation}
B(p)=\left[\begin{array}{cccccc}
p & 1-p & 0 & 0 & \cdots & 0\\
0 & 0 & p & 0 & \cdots & 0\\
0 & 0 & 0 & p & \cdots & 0\\
\vdots &  \vdots & \vdots & \vdots & \ddots & \vdots\\
0 & 0 & 0 & 0 & \cdots & p\\
p & 0 & 0 & 0 & \cdots & 0\end{array}\right].\label{statematrix}\end{equation}

\end{lemma}
\begin{IEEEproof} See Appendix~\ref{app_lemma53}. \end{IEEEproof}

Let us look at (\ref{ub_unver})
, we can see that the average number of unverified symbols scales exponentially with $k$.  The ensemble with larger $\lambda_{2}\rho^{'}(1)$ will have
more short degree-2 cycles and more average unverified symbols. The average number of unverified symbols depends on the maximum list-size $S_{max}$ in a complicated way. Intuitively, 
if $S_{max}$ is larger, then the constraint that "there is at most one correct symbol along $S_{max}$ adjacent variable nodes`` becomes stronger since we assume the probability of 
seeing a correct symbol is higher than that of seeing a incorrect symbol. Therefore, unverification is  less likely to happen and the average number of unverified symbols will decrease as $S_{max}$ increases. 
Notice also that (\ref{ub_unver}) does not depend on $q$.


%

\begin{small} %
\begin{table*}[!ht]

\caption{Optimization Results for LMP Algorithms (rate 1/2)}

\label{tab:opt_our} \centering \begin{tabular}{|c|c|c|c|}
\hline 
\textbf{Alg.}  & $\boldsymbol{\lambda}\mathbf{(x)}$  & $\boldsymbol{\rho}\mathbf{(x)}$  & $\mathbf{p^{*}}$ \tabularnewline
\hline  LMP-1  & $.1200x\!+\!.3500x^{2}\!+\!.0400x^{4}\!+\!.4900x^{14}$ & $x^{8}$ & .2591 \tabularnewline
\hline LMP-1  & $.1650x\!+\!.3145x^{2}\!+\!.0085x^{4}\!+\!.2111x^{14}\!+\!.0265x^{24}\!+\!.0070x^{34}\!+\!.2674x^{49}$ & $.0030x^{2}\!+\!.9970x^{10}$ & .2593 \tabularnewline
\hline 
LMP-8 & $.32x\!+\!.24x^{2}\!+\!.26x^{8}\!+\!.19x^{14}$ & $.02x^{4}\!+\!.82x^{6}\!+\!.16x^{8}$ & .288  
\tabularnewline
\hline 
LMP-32 & $.40x\!+\!.20x^{3}\!+\!.13x^{5}\!+\!.04x^{8}\!+\!.23x^{14}$ & $.04x^{4}\!+\!.96x^{6}$ & .303 \tabularnewline
\hline 
LMP-$\infty$ & $.34x\!+\!.16x^{2}\!+\!.21x^{4}\!+\!.29x^{14}$  & $x^{7}$  & .480 \tabularnewline
\hline 
LM2-MB & $.2x\!+\!.3x^{3}\!+\!.05x^{5}\!+\!.45x^{11}$  & $x^{8}$  & .289 \tabularnewline
\hline 
LM2-NB & $0.2972x\!+\!0.1560x^{2}\!+\!0.1035x^{3}\!+\!0.1813x^{5}\!+\!0.2590x^{8}\!+\!0.0029x^{11} $  & $x^{6}$  & .303 \tabularnewline
\hline
\end{tabular}
\end{table*}

\end{small}

One might expect that the stability condition of the LMP-$S_{max}$ decoding algorithms 
can be used to analyze the error floor. Actually, one can show that the stability condition for LMP-$S_{max}$ decoding
of irregular LDPC codes is identical to that of the BEC, which is 
$p \lambda_2 \rho'(1) < 1$.
This does not help predict the error floor though, because
for codes with degree-2 nodes, the error floor is determined
mainly by short cycles of degree-2 nodes.
Instead, the condition $\lambda_2 \rho'(1) < 1$ simply implies that
the expected number of degree-2 cycles is finite.

\section{Comparison and Optimization}
\label{sec:cmp_opt}

In this section, we compare the proposed algorithm, with maximum list
size $S_{max}$ (LMP-$S_{max}$), to other message-passing decoding algorithms
for the $q$-SC. We note that the LM2-MB algorithm is identical to
SW1 for any code ensemble because the decoding rules are the same.
LM2-MB, SW1 and LMP-1 are identical for (3,6) regular LDPC codes because
the list size is always 1 and erasures never occur in LMP-1 for (3,6)
regular LDPC codes. The LMP-$\infty$ algorithm is identical to
SW2.

There are two important differences between the LMP algorithm and
previous algorithms: (i) erasures and (ii) FV recovery. The LMP algorithm
passes erasures because, with a limited list size, it is better to
pass an erasure than to keep unlikely symbols on the list. The LMP
algorithm also detects FV events and passes an erasure if they cause
disagreement between verified symbols later in decoding, and can sometimes
recover from a FV event. In contrast, LM1-NB and LM2-NB fix the status of a variable
node once it is verified and pass the verified value in all following
iterations.

The results in \cite{my_ref:luby_and_mitz} and \cite{my_ref:Shokwangpersonal}
also do not consider the effects of type-II FV. These FV events degrade
the performance in practical systems with moderate block lengths, and
therefore we use random entries in the parity-check matrix to mitigate
these effects.

Using the DE analysis of the LMP-$S_{max}$ algorithm, we can improve the
threshold by optimizing the degree distribution pair $(\lambda,\rho)$.
Since the DE recursion is not one-dimensional, we use differential
evolution to optimize the code ensembles \cite{Storn-jgo97}. In Table~\ref{tab:opt_our},
we show the results of optimizing rate-$\frac{1}{2}$ ensembles for
LMP with a maximum list size of 1, 8, 32, and $\infty$. Thresholds
for LM2-MB and LM2-NB algorithms with rate 1/2 are also shown. In all but one
case, the maximum variable-node degree is 15 and the maximum check-node
degree is 9.  The second table
entry allowed for larger degrees (in order to improve performance)
but very little gain was observed. We can also see that there is a
gain of between 0.04 and 0.07 over the thresholds of (3,6) regular
ensemble with the same decoder.

\section{Simulation Results}
\label{sec:sim}

\begin{figure}[t]
  
\centering
\includegraphics[width=0.35\columnwidth,viewport=220 30 450 500]{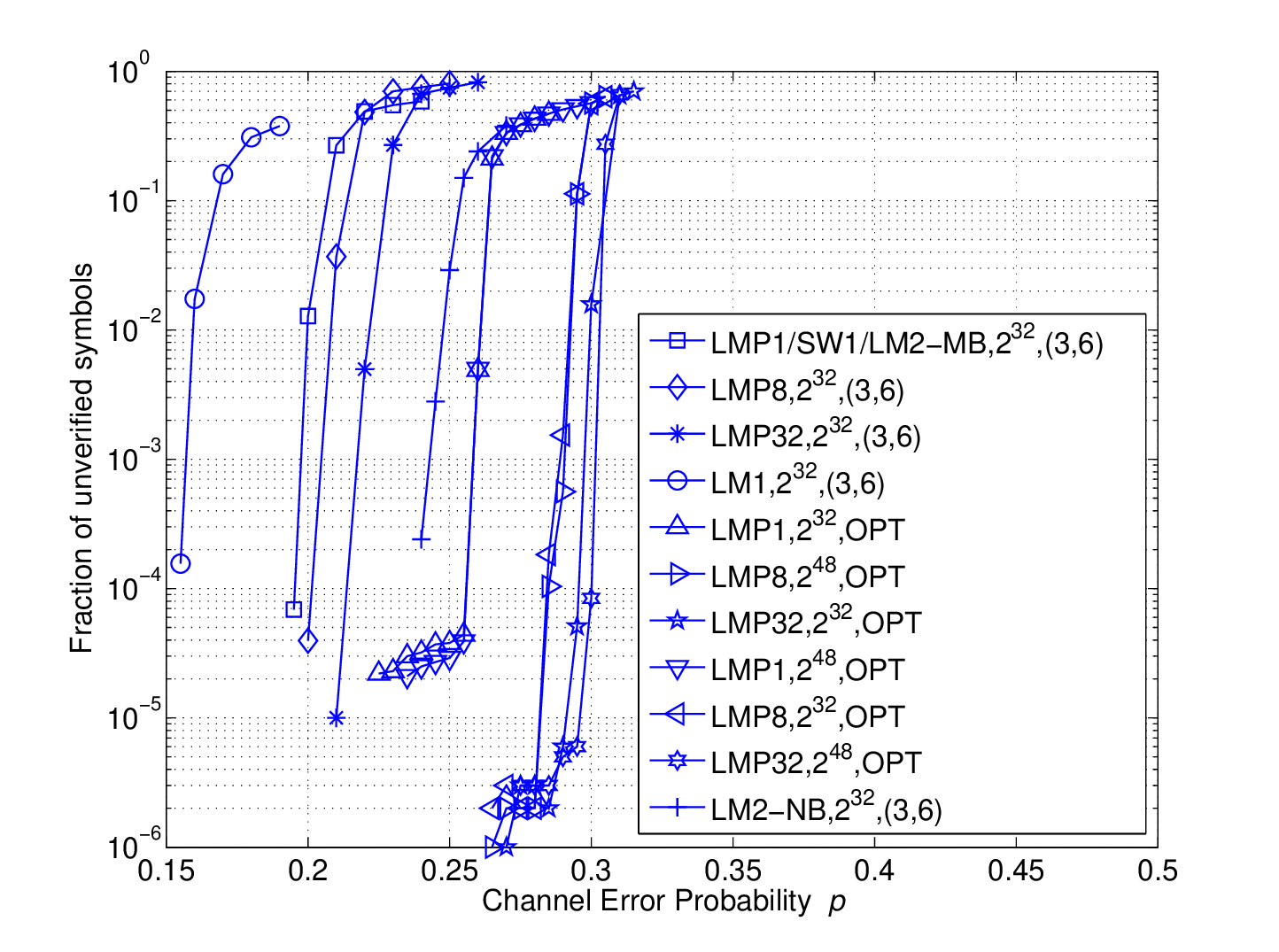}
\caption{Simulation results for (3,6) regular codes with block
length 100000.}\label{fig:sim36}
\end{figure}

In this section, simulation results are presented for regular and optimized LDPC
codes (see Table~\ref{tab:opt_our}) using various decoding algorithms and maximum list sizes.
For the simulation of optimized ensembles, a variety of finite-field sizes are also tested.
We use notation {}``LMP$S_{max}$,$q$,$X$'' to denote the simulation results
of the LMP algorithm with maximum list-size $S_{max}$, finite field $GF(q)$, and ensemble $X$.
The block length is chosen to be 100000 and the parity-check matrices are chosen randomly while avoiding double edges and 4-cycles.
Each non-zero entry in the parity-check matrix is chosen uniformly
from $\textrm{GF}(q)\setminus0$ to minimize the FV probability.
The maximum number of decoding iterations is fixed to be 200
and more than 1000 blocks are run for each point.
The results are shown in Fig.~\ref{fig:sim36} and can be compared with the theoretical thresholds.
Table~\ref{tab:thresh36} shows the theoretical thresholds of $(3,6)$ regular codes on the
$q$-SC for different algorithms and Table~\ref{tab:opt_our} shows the thresholds
for the optimized ensembles.
The numerical results seem to match the theoretical thresholds well.

\begin{small} %
\begin{table}[t]

\caption{Threshold vs. algorithm for the (3,6) regular LDPC ensemble}

\renewcommand{\tabcolsep}{0.15cm}
\label{tab:thresh36} \centering \begin{tabular}{@{}|c|c|c|c|c|c|c|c|@{}}
\hline 
LMP-1 & LMP-8 & LMP-32 & LMP-$\infty$ & LM1 & LM2-MB & LM2-NB  \tabularnewline
\hline 
.210 & .217 & .232 & .429 & .169 & .210 & .259  \tabularnewline
\hline
\end{tabular}
\end{table}

\end{small}


The results of the simulation for (3,6) regular codes shows no apparent
error floor because there are no degree-2 nodes and almost no FV occurs is observed during simulation.
The LM2-NB performs much better than other algorithms with list-size 1 for the (3,6)
regular ensemble. For the optimized ensembles, there are a large number of degree-2
variable nodes which cause a significant error floor.
By evaluating (\ref{ub_unver}), the predicted
error floor caused by unverification is $1.6\times10^{-5}$ for the
optimized $S_{max}=1$ ensemble, $8.3\times10^{-7}$ for the optimized
$S_{max}=8$ ensemble, and $1.5\times10^{-6}$ for the optimized $S_{max}=32$
ensemble. From the results, we see the analysis of unverification
events matches the numerical results very well.


\section{Conclusions}
\label{sec:con}
In this paper, list-message-passing (LMP) decoding algorithms are discussed
for the $q$-ary symmetric channel ($q$-SC). It is shown that capacity-achieving
ensembles for the BEC achieve capacity on the $q$-SC when the list
size is unbounded and $q$ goes to infinity. Decoding thresholds are also calculated by density
evolution (DE). We also derive a new analysis for the node-based algorithms
described in \cite{my_ref:luby_and_mitz}. The causes of false verification
(FV) are also analyzed and random entries in the parity-check matrix are
used to avoid type-II FV.  Degree profiles are optimized for
the LMP decoder and reasonable gains are obtained. Finally, simulations
show that, with list sizes larger than 8, the proposed LMP algorithm
outperforms previously proposed algorithms. In the simulation, we also observe significant error floors
for the optimized code ensembles. These error floors are mainly caused by the unverified symbols when decoding terminates.
An analysis of this error floor is derived that matches the simulation results quite well. 

While we focus on the $q$-SC in this work, there are a number of
other applications of LMP decoding that are also quite interesting.
For example, the iterative decoding algorithm described in \cite{my_ref:isit2006sarvotham}
for compressed sensing is actually the natural extension of LM1 to
continuous alphabets. For this reason, the LMP decoder may also be
used to improve the threshold of compressed sensing. This is, in some
sense, more valuable because there are a number of good coding schemes
for the $q$-SC, but few low-complexity near-optimal decoders for
compressed sensing. This extension is explored more thoroughly in \cite{zp-it-cs}.

\appendices{ }

\begin{section}{Proof of theorem \ref{thm1}}
\label{app_thm1}


\begin{IEEEproof} Given $p\lambda(1-\rho(1-x))<x$ for $x\in(0,1]$,
we start by showing that both $x_{i}$ and $y_{i}$ go to zero as
$i$ goes to infinity. To do this, we let $\alpha=\sup_{x\in(0,1)}\frac{1}{x}p\lambda(1-\rho(1-x))$
and note that $\alpha<1$ because $p<p^{*}$. It is also easy to see
that, starting from $x_{0}=1$, we have $x_{i}\leq\alpha^{i}$ and
$x_{i}\rightarrow0$. Next, we rewrite (\ref{eq3}) as \begin{align*}
y_{i+1} & =\frac{1}{p}x_{i+1}+p\left(\rho(1-x_{i})-\rho(1-y_{i})\right)\lambda'(1-\rho(1-x_{i}))\\
 & \stackrel{(a)}{\leq}\frac{1}{p}\alpha^{i+1}+p\left(1-\rho'(1)\alpha^{i}-\rho(1-y_{i})\right)\left(\lambda_{2}+O(\alpha^{i})\right)\\
 & \stackrel{(b)}{\leq}\frac{1}{p}\alpha^{i+1}+p\lambda(1-\rho(1-y_{i}))\left(1+O(\alpha^{i})\right)\\
 & \stackrel{(c)}{\leq}\frac{1}{p}\alpha^{i+1}+\alpha y_{i}\left(1+O(\alpha^{i})\right),\end{align*}
 where $(a)$ follows from $\rho(1-x)\leq1-\rho'(1)x$, $(b)$ follows
from $\lambda_{2}(1-\rho(1-y))\leq\lambda(1-\rho(1-y))$, and $(c)$
follows from $p\lambda(1-\rho(1-y))\leq\alpha y$. It is easy to verify
that $y_{i+1}<y_{i}$ as long as $y_{i}>\frac{\alpha^{i+1}}{p(1-\alpha(1+O(\alpha^{i})))}$.
Therefore, we find that $y_{i}\rightarrow0$ because the recursion
does not have any positive fixed points as $i\rightarrow\infty$.
Moreover, one can show that $y_{i}$ eventually decreases exponentially
at a rate arbitrarily close to $\alpha$.


Now, we consider the performance of a randomly chosen code and with a random error pattern.
The approach taken is general enough to cover all the message-based decoding algorithms discussed in this paper.

A message in the decoding graph is called \emph{bad} if it is either unverified
or falsely verified.  Based on \cite{my_ref:DE}, one can show that, after $\ell$ decoding iterations,
the fraction of bad messages is tightly concentrated around its average.
We note that the concentration occurs regardless of whether there are falsely verified messages or short cycles.
Since the proof involves no new techniques beyond \cite{my_ref:DE} and the extension to irregular codes in \cite{my_ref:kavcic-capacity}, we give only a brief outline.
The main difference between this scenario and \cite{my_ref:DE} is that the algorithms discussed in this paper pass lists whose size may be unbounded and may be affected by FV.
It turns out that the size of these lists is superfluous, however, if we only consider the fraction of messages satisfying a logical test (e.g., the fraction of bad messages).

Let the r.v. $Z^{(\ell)}$ denote the number of \emph{bad} variable-to-check messages after $\ell$ iterations of decoding.
Suppose the fraction of unverified messages predicted by DE, assuming no FV, is $y_{\ell}$.
We say that concentration fails if $Z^{(\ell)}/E$ exceeds $y_{\ell}$ by more than $\epsilon$, where $E$ is the number of edges in the graph. 
Following \cite{my_ref:DE}, one can bound the failure probability by showing that the fraction of bad messages concentrates around its average value and that the average value converges to the value computed by DE.

The main observation is that the verification status of an edge, after $\ell$ iterations, depends only on the depth-$2\ell$ neighborhood, $\mathcal{N}_{\vec{e}}^{(2\ell)}$, of the directed edge $\vec{e}$.
By using a Doob's martingale for the edge-exposure process and applying Azuma's inequality, one obtains the concentration bound
\begin{equation}
\label{eq:Doob}
\Pr\left(\frac{Z^{(\ell)}}{E}-\frac{\mathbb{E}\left[Z^{(\ell)}\right]}{E} > \frac{\epsilon}{3}\right) \le e^{-\beta_{\ell} \epsilon^{2}n},
\end{equation}
where $\beta_{\ell}$ is a positive constant independent of $n$ which depends only on the d.d. and the number of iterations.



The next step is showing that the expected value $\mathbb{E}\left[Z^{(\ell)}\right]/E$
is close to the value, $y_{\ell}$, given by DE.  In this case, we must consider
two sources of error: short cycles and false verification.
In \cite{my_ref:DE} and \cite{my_ref:kavcic-capacity}, it is shown
that, when a code graph is chosen uniformly at random from all possible
graphs with degree distribution pair $(\lambda(x),\rho(x))$,
\[ \Pr\left(\mathcal{N}_{\vec{e}}^{(2\ell)}\mbox{ is not tree-like}\right)\le\frac{\gamma_{\ell}}{n},\]
 where $\gamma_{\ell}$ is a constant independent of $n$ that depends only on the d.d. and number of iterations.
So, for any $\epsilon>0$,
one can choose $n$ large enough so that $\gamma_{\ell} /n < \epsilon/3$.
Since type-II FV is a failure due to short cycles
(see  Section \ref{false_v} for details about type-I and type-II FV),
this implies that any increase in 
$\mathbb{E}\left[Z^{(\ell)}\right]/E$ due to short cycles and type-II FV is at most $\epsilon/3$.
Likewise, one can upper bound the effect of type-I FV with
\[ \Pr\left(\mbox{any type-I FV occurred in }\mathcal{N}_{\vec{e}}^{(2\ell)}\right)\le\frac{\theta_{\ell} }{q},\]
where $\theta_{\ell}$ is a constant independent of $n$ that depends only the d.d., the number of iterations, and algorithm details (e.g., $S_{max}$).
If we choose $q$ large enough so that $\theta_{\ell} /q < \epsilon/3$, then
these two bounds imply that
\begin{equation}
\label{eq:mean_shift}
\frac{\mathbb{E}\left[Z^{(\ell)}\right]}{E} - y_{\ell} \leq \frac{\gamma_{\ell}}{n} + \frac{\theta_{\ell}}{q} \leq \frac{2\epsilon}{3}.
\end{equation}
Finally, \eqref{eq:Doob} and \eqref{eq:mean_shift} can be combined to bound the probability that
$Z^{(\ell)}/E$ is greater than $y_{\ell}+\epsilon$.
\end{IEEEproof}
\end{section}

\begin{section}{Proof of Lemma \ref{lemma3}}
\label{app_lemma3}
Let $y$ be the received symbol sequence assuming the
all-zero codeword was transmitted and let $u$ be another codeword with exactly
$k$ non-zero symbols.
Of the $k$ positions where they differ, assume that $i$ are received correctly, $j$ are flipped to other codeword's value, and $k-i-j$ are flipped to a third value.
It is easy to verify that the all-zero codeword is not the unique ML codeword whenever $i\leq j$.
Therefore, the probability that ML decoder chooses $u$ over the all-zero codeword is given by
\begin{equation*}
p_{2,k} =\sum_{j=0}^{k}\sum_{i=0}^{j}\textstyle \binom{k}{i,j,k-i-j} (1\!-\!p)^{i}\left(\frac{p}{q-1}\right)^{j}\left(\frac{p(q-2)}{q-1}\right)^{k-i-j}.\end{equation*}
The multinomial theorem also shows that
\begin{align*}
A& (x) = \textstyle \left((1\!-\!p)+\frac{p}{q-1}x^{2}+\frac{p(q-2)}{q-1}x\right)^{k} \\
& = \displaystyle \sum_{j=0}^{k}\sum_{i=0}^{k-j}\textstyle \binom{k}{i,j,k-i-j} (1\!-\!p)^{i} \! \left(\frac{p}{q-1}\right)^{j} \! \left(\frac{p(q-2)}{q-1}\right)^{k-i-j} \!\! x^{k-i+j} \\ & \triangleq \displaystyle \sum_{l=0}^{2k} A_{l} x^{l},
\end{align*}
where $A_{l}$ is the coefficient of $x^l$ in $A(x)$.
Next, we observe that $p_{2,k} = \sum_{l=k}^{2k} A_{l}$ is simply an
unweighted sum of a subset of terms in $A(x)$ (namely, those where $k-i+j \geq k$).

This implies that
\[ x^{k}p_{2,k}=\sum_{l=k}^{2k}A_{l} x^{k} \le A(x)\]
 for any $x\ge1.$ Therefore, we can compute the Chernoff-type bound
\[ p_{2,k}\le\inf_{x\ge1}x^{-k}A(x).\]
 By taking derivative of $x^{-k}A(x)$ over $x$ and setting it to
zero, we arrive at the bound
\[p_{2,k} \leq \left(p\frac{q-2}{q-1} +\sqrt{\frac{4p(1-p)}{q-1}} \right)^k.\]

 \end{section}

\begin{section}{Proof of Lemma \ref{lemma53}}
\label{app_lemma53}
\begin{figure}[t]
\begin{center}
\includegraphics[width=0.4\columnwidth]{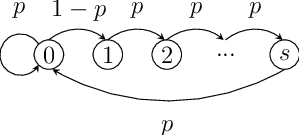}
\caption{Finite-state machine for Lemma \ref{lemma53}.}
\end{center}
\label{fig:fsm} 
\end{figure}

\begin{IEEEproof}  An unverification event occurs on a degree-2 cycle
of length-$k$ when there is at most one correct variable node in any
adjacent set of $s+1$ nodes.
Let the set of all error patterns (i.e., 0 means correct and 1 means error)
of length-$k$ which satisfy the UV condition be $\Phi(s,p,k)\subseteq \{0,1\}^k$.
Using the Hamming weight $w(z)$, of an error pattern as $z$, to count the number of
errors, we can write the probability of UV  as
\[ \phi(s,p,k)=\sum_{z\in\Phi(s,p,k)}p^{w(z)}(1-p)^{k-w(z)}.\]

 This expression can be evaluated using the transfer matrix method to
enumerate all weighted walks through a particular digraph. If we walk through the nodes along
the cycle by picking an arbitrary node as the starting node, the UV
constraint can be seen as $k$-steps of a particular finite-state machine.
Since we are walking on a cycle, the initial state must equal to the final
state.

The finite-state machine, which is shown in Fig. \ref{fig:fsm}, has $s+1$ states $\{0,1,\ldots,s\}$.
Let state 0 be the state where we are free to choose either a correct or incorrect symbol
(i.e., the previous $s$ symbols are all incorrect).
This state has a self-loop associated with the next symbol also being incorrect.
Let state $i>0$ be the state where the past $i$ values consist of one correct symbol followed by
$i-1$ incorrect symbols.
Notice that only state 0 may generate correct symbols.
By defining the transfer matrix with (\ref{statematrix}), the
probability that the UV condition holds is therefore $\phi(s,p,k)=\textrm{Tr}\left(B^{k}(p)\right)$.
\end{IEEEproof}

 \end{section}

\bibliographystyle{ieeetr}

\begin{thebibliography}{10}

\bibitem{my_ref:r1}
R.~G. Gallager, ``Low-density parity-check codes,'' {\em IRE Trans. Inform.
  Theory}, vol.~18, pp.~21--28, Jan. 1962.

\bibitem{mackay1995good}
D.~MacKay and R.~Neal, ``{Good codes based on very sparse matrices},'' {\em
  Lecture Notes in Computer Science}, vol.~1025, pp.~100--111, 1995.

\bibitem{my_ref:lubyca1}
M.~Luby, M.~Mitzenmacher, M.~Shokrollahi, and D.~Spielman, ``Efficient erasure
  correcting codes,'' {\em {IEEE} Trans. Inf. Theory}, vol.~47, pp.~569--584,
  Feb. 2001.

\bibitem{my_ref:urbanke}
T.~Richardson, M.~Shokrollahi, and R.~Urbanke, ``Design of capacity-approaching
  irregular low-density parity-check codes,'' {\em {IEEE} Trans. Inf. Theory},
  vol.~47, pp.~619--637, Feb. 2001.

\bibitem{my_ref:ShokITW04}
A.~Shokrollahi, ``Capacity-approaching codes on the $q$-ary symmetric channel
  for large $q$,'' in {\em Proc. IEEE Inform. Theory Workshop}, (San Antonio,
  TX), pp.~204--208, Oct. 2004.

\bibitem{my_ref:isit2008weidmann}
C.~Weidmann, ``Coding for the $q$-ary symmetric channel with moderate $q$,'' in
  {\em Proc. IEEE Int. Symp. Information Theory}, July 2008.

\bibitem{my_ref:LW_turbo08}
G.~Lechner and C.~Weidmann, ``Optimization of binary ldpc codes for the $q$-ary
  symmetric channel with moderate $q$,'' in {\em Proc. 5th International
  Symposium on Turbo Codes and Related Topics}, 2008.

\bibitem{my_ref:luby_and_mitz}
M.~Luby and M.~Mitzenmacher, ``Verification-based decoding for packet-based
  low-density parity-check codes,'' {\em {IEEE} Trans. Inf. Theory}, vol.~51,
  pp.~120--127, Jan. 2005.

\bibitem{my_ref:Metz}
J.~Metzner, ``Majority-logic-like decoding of vector symbols,'' {\em {IEEE}
  Trans. Commun.}, vol.~44, pp.~1227--1230, Oct. 1996.

\bibitem{my_ref:Metz2}
J.~Metzner, ``Majority-logic-like vector symbol decoding with alternative
  symbol value lists,'' {\em {IEEE} Trans. Commun.}, vol.~48, pp.~2005--2013,
  Dec. 2000.

\bibitem{my_ref:Davey}
M.~Davey and D.~MacKay, ``Low density parity check codes over {GF}($q$),'' {\em
  {IEEE} Commun. Lett.}, vol.~2, pp.~58--60, 1998.

\bibitem{my_ref:BKY03}
D.~Bleichenbacher, A.~Kiyayias, and M.~Yung, ``Decoding of interleaved
  {R}eed-{S}olomon codes over noisy data,'' in {\em Proc. of ICALP},
  pp.~97--108, 2003.

\bibitem{my_ref:isit2004wangshok}
A.~Shokrollahi and W.~Wang, ``Low-density parity-check codes with rates very
  close to the capacity of the $q$-ary symmetric channel for large $q$,'' in
  {\em Proc. IEEE Int. Symp. Information Theory}, (Chicago, IL), p.~275, June
  2004.

\bibitem{my_ref:Shokwangpersonal}
A.~Shokrollahi and W.~Wang, ``Low-density parity-check codes with rates very
  close to the capacity of the $q$-ary symmetric channel for large $q$.''
  Unpublished extended abstract, 2004.

\bibitem{my_ref:DE}
T.~Richardson and R.~Urbanke, ``The capacity of low-density parity-check codes
  under message-passing decoding,'' {\em {IEEE} Trans. Inf. Theory}, vol.~47,
  pp.~599--618, Feb. 2001.

\bibitem{my_ref:Guruswami03}
V.~Guruswami and P.~Indyk, ``Linear time encodable and list decodable codes,''
  in {\em Proc.\ of the 35th Annual ACM Symp.\ on Theory of Comp.},
  pp.~126--135, 2003.

\bibitem{Zhang-unpub11}
F.~Zhang and H.~D. Pfister, ``On the stopping sets of verification decoding.''
  in preparation, June 2011.

\bibitem{my_ref:diffeqn}
N.~Wormald, ``Differential equations for random processes and random graphs,''
  {\em Annals of Applied Probability}, vol.~5, pp.~1217--1235, 1995.

\bibitem{Zhang-web09}
F.~Zhang and H.~D. Pfister, ``Software to compute the {LM2-NB} threshold.'' at
  http://www.ece.tamu.edu/\textasciitilde
  hpfister/software/lm2nb\textunderscore threshold.m.

\bibitem{wormald_de_1997}
N.~C. Wormald, ``{The differential equation method for random graph processes
  and greedy algorithms},'' in {\em Lectures on Approximation and Randomized
  Algorithms}, pp.~73--155, Polish Scientific Publishers, 1999.

\bibitem{RU-2008}
T.~J. Richardson and R.~L. Urbanke, {\em Modern Coding Theory}.
\newblock Cambridge University Press., 2008.

\bibitem{Storn-jgo97}
R.~Storn and K.~Price, ``Differential evolution--{A} simple and efficient
  heuristic for global optimization over continuous spaces,'' {\em J. Global
  Optim.}, vol.~11, no.~4, pp.~341--359, 1997.

\bibitem{my_ref:isit2006sarvotham}
S.~Sarvotham, D.~Baron, and R.~Baraniuk, ``Sudocodes--{F}ast measurement and
  reconstruction of sparse signals,'' in {\em Proc. IEEE Int. Symp. Information
  Theory}, (Seattle, WA), pp.~2804--2808, July 2006.

\bibitem{zp-it-cs}
F.~Zhang and H.~D. Pfister, ``Veri�cation decoding of high-rate ldpc codes with
  applications in compressed sensing.'' submitted to {\em IEEE Trans. on
  Inform. Theory} also available in Arxiv preprint cs.IT/0903.2232v3, 2009.

\bibitem{my_ref:kavcic-capacity}
A.~Kavcic, X.~Ma, and M.~Mitzenmacher, ``Binary intersymbol interference
  channels: Gallager codes, density evolution, and code performance bounds,''
  {\em {IEEE} Trans. Inf. Theory}, vol.~49, pp.~1636--1652, July 2003.

\end{thebibliography}

\begin{IEEEbiographynophoto}{Fan Zhang} (S'03)
received his Ph.D. in electrical engineering from Texas A\&M University in 2010 and joined the read-channel arthitecture group at LSI Logic. He spent 3 years at UTStarcom Inc. before joining Texas A\&M University.

His current research interests include information theory, error correcting codes and signal processing for wireless and data storage systems.
\end{IEEEbiographynophoto}
\begin{IEEEbiographynophoto}{Henry D. Pfister} (S'99--M'03--SM'09)
 received his Ph.D. in electrical engineering from UCSD
in 2003 and he joined the faculty of the School of Engineering at Texas A\&M
University in 2006.  Prior to that he spent two years in R\&D at
Qualcomm, Inc.
and one year as a post-doc at EPFL.

He received the NSF Career Award in 2008 and was a coauthor of the 2007
IEEE COMSOC best paper in Signal Processing and Coding for Data
Storage.

His current research interests include information theory, channel coding,
and iterative decoding with applications in wireless communications and
data storage.
\end{IEEEbiographynophoto}

\end{document}